\documentclass[english,11pt,a4paper]{article}
\usepackage[T1]{fontenc}
\usepackage[latin9]{inputenc}
\usepackage{amsmath}
\usepackage{amsthm}
\usepackage{amssymb}
\usepackage{graphicx}

\makeatletter
\numberwithin{equation}{section}

\usepackage{jheppub}
\usepackage{orcidlink}

\usepackage{xcolor}

\makeatother

\usepackage{babel}
\begin{document}
\title{High-Energy and Ultra-High-Energy  Neutrinos from  Primordial
Black Holes}

\abstract{ Primordial Black Holes (PBHs) are capable of emitting
extremely energetic particles independent of their interactions with
the Standard Model. In this work, we investigate whether PBHs evaporating
in the early universe could be responsible for some of the observed
high-energy neutrinos above the TeV or PeV scale in the present universe.
We compute the energy spectrum of neutrinos directly emitted by PBHs
with a monochromatic mass function and estimate the wash-out point,
which determines the maximum energy of the spectrum. We find that
the spectrum generally  extends to high energies following a power
law of $E_{\nu}^{-3}$ until it reaches the wash-out point, which
crucially depends on the PBH mass. For PBHs of $10^{13}$ grams, the
spectrum can extend up to the PeV scale, though the flux is too low
for detection. We also consider an indirect production mechanism involving
dark particles that are emitted by PBHs and decay into neutrinos at
a much later epoch. This mechanism allows lighter (such as those in
the gram to kilogram range) PBHs to produce more energetic neutrino
fluxes without being washed out by the thermal plasma in the early
universe. In this scenario, we find that ultra-high-energy neutrinos
around or above the EeV scale can be generated, with sufficiently
high fluxes detectable by current and future high-energy neutrino
observatories  such as IceCube and GRAND.}

\author[a]{Quan-feng Wu \orcidlink{0000-0002-5716-5266}}
\author[a]{and Xun-Jie Xu \orcidlink{0000-0003-3181-1386}}
\affiliation[a]{Institute of High Energy Physics, Chinese Academy of Sciences, Beijing 100049, China}
\emailAdd{wuquanfeng@ihep.ac.cn} 
\emailAdd{xuxj@ihep.ac.cn} 
\preprint{\today}  
\maketitle

\section{Introduction}

One of the most compelling questions in our quest to understand the
universe is the origin of cosmic rays with extremely high energies.
The most energetic cosmic ray events observed exceed hundreds of EeV,
which is orders of magnitude higher than the energy attainable at
colliders. Various explanations for the presence of such energetic
particles have been proposed, including supernova remnant shocks via
Fermi acceleration~\cite{Fermi:1949ee}, active galactic nuclei (AGNs)
\cite{Blandford:2018iot},  dark matter annihilation~\cite{Blasi:2001hr},
and the decays of supermassive particles~\cite{Hill:1982iq,Birkel:1998nx,Kuzmin:1998uv}\,---\,see
e.g.~\cite{Bhattacharjee:1999mup,Torres:2004hk,Anchordoqui:2018qom,Ackermann:2022rqc}
for reviews. 

In this work, we propose an interesting origin of ultra-high-energy
cosmic rays: primordial black holes (PBHs)~\cite{Carr:2009jm,Carr:2020gox}.
According to Hawking radiation, a black hole with a sufficiently low
mass emits high-energy radiation with the energy comparable to the
black hole temperature. Since the temperature is proportional to the
inverse of the black hole mass which decreases during evaporation,
one expects that particles emitted at the very final stage of the
evaporation can be extremely energetic, possibly reaching the Planck
scale. Therefore, PBHs with lifetimes shorter than the age of the
universe may have left high-energy radiations in the early universe
and could contribute significantly to the high-energy cosmic rays
observed today. 

Given that the early universe is filled with a hot, dense plasma,
most of the particles emitted by PBHs would lose energy rapidly 
as they propagate through the plasma, unless their interactions with
the plasma are sufficiently weak. Neutrinos, however,   due to their
weakly-interacting nature may be able to retain the energetic form
originating from PBHs under certain conditions. Therefore, the primary
goal of this study is to explore the possibility of generating high-energy
neutrinos from  PBH evaporation in the early universe. 

We note here that the idea of generating neutrino fluxes from PBHs
has been previously investigated in Ref.~\cite{Lunardini:2019zob},
though with a different focus on the Dirac and Majorana nature of
neutrinos.  In Ref.~\cite{Lunardini:2019zob},  the so-called diffuse
neutrino flux from PBHs has been computed, but the obtained spectrum
did not extend to high energies above GeV. We will show that neutrinos
from  PBHs  may feature a hard tail in the spectrum without exponential
suppression, allowing their energies to extend up to the PeV scale
for $10^{13}$-gram PBHs. Heavier PBHs in principle could produce
more extended spectra. However, PBHs above $10^{15}$ grams cannot
fully evaporate within time intervals shorter than the age of the
universe  and hence cannot generate the aforementioned hard tail.\footnote{Nevertheless, their evaporation in the present universe still generates
neutrino fluxes within an interesting energy range (typicallly around
$10$ to $30$ MeV)   which could be observable in next-generation
neutrino detectors~\cite{Halzen:1995hu,Bugaev:2000bz,Bugaev:2002yt,Wang:2020uvi,DeRomeri:2021xgy,Calabrese:2021zfq,Bernal:2022swt,Liu:2023cqs}. } 

In addition to the production of high-energy neutrinos directly from
PBHs, we also consider an indirect production mechanism involving
heavy dark particles that are emitted by PBHs and later decay into
neutrinos. The dark particles could be heavy neutral leptons~\cite{DeRomeri:2024zqs}
but in this work are very generic. This mechanism utilizes another
prominent feature of PBHs: heavy dark particles can be effectively
produced from PBHs, independent of their interactions with the Standard
Model. This feature has been extensively exploited in the literature
to address the dark matter and baryon asymmetry problems~\cite{Baldes:2020nuv,Gondolo:2020uqv,Bernal:2020kse,Bernal:2020bjf,Bernal:2020ili,Auffinger:2020afu,Hooper:2020otu,Datta:2020bht,Cheek:2021odj,Masina:2021zpu,Cheek:2021cfe,Sandick:2021gew,Bernal:2021yyb,Bernal:2021bbv,Calabrese:2021src,JyotiDas:2021shi,Bernal:2022pue,Bernal:2022oha,Coleppa:2022pnf,Gehrman:2022imk,Calabrese:2023key,Schmitz:2023pfy,Gehrman:2023esa,Gehrman:2023qjn,DeRomeri:2024zqs,Arcadi:2024tib}.
 In our work, introducing heavy dark particles allows us to make
use of very small PBHs, that evaporate well before neutrino coupling,
to still produce high-energy neutrino fluxes without being washed
out by the dense thermal plasma. We show that with this indirect production
mechanism, PBHs in the gram to kilogram range can generate ultra-high-energy
neutrinos around or above the EeV scale, with sufficiently high fluxes
detectable by current and future high-energy neutrino observatories
such as IceCube~\cite{IceCube-Gen2:2020qha} and GRAND~\cite{GRAND:2018iaj}.

This paper is organized as follows. In Sec.~\ref{sec:The-hard-spectrum},
we  calculate the cumulative energy spectrum of particles radiated
by a PBH evaporating in static spacetime, and show that after full
evaporation this spectrum features a hard tail which is not exponentially
suppressed.  In Sec.~\ref{sec:High-E-in-universe}, we consider
PBHs evaporating in the expanding universe, taking into account cosmological
redshift in the calculation of the energy spectrum. Various cases,
including massless, massive stable, and slow-decaying species, are
discussed and calculated. In Sec.~\ref{sec:High-energy-neutrinos},
we apply our calculations to neutrinos and present the obtained neutrino
fluxes for a few benchmarks. Finally, we conclude in Sec.~\ref{sec:Conclusion}
and relegate some details to the appendices.

\section{The hard spectrum \label{sec:The-hard-spectrum} }

One of the most intriguing features of PBHs is that, as they evaporate,
their temperatures keep increasing, possibly up to the Planck scale.
As a consequence, at the very final stage of the evaporation, they
are able to emit extremely energetic radiations or very heavy particles,
independent of their interactions with the SM. 

The instantaneous energy spectrum of a generic particle species emitted
by an evaporating PBH is close to the Fermi-Dirac or Bose-Einstein
distributions with exponential suppression at energies well above
the PBH temperature. If the emitted particle is stable or long-lived,
the cumulative energy spectrum after evaporation possesses a hard
tail which follows a power law instead of exponential suppression. 

More specifically, let us consider the Hawking radiation rate~\cite{Page:1976df}\footnote{See also Refs.~\cite{Baldes:2020nuv,Cheek:2021odj} for more recent
 discussions on this  rate.}:
\begin{align}
\frac{d^{2}N_{i}}{dtdE} & \approx\frac{g_{i}}{2\pi}\cdot\frac{\gamma_{\text{gray}}}{\exp(E/T_{{\rm BH}})-\eta}\thinspace,\label{eq:PBH-1}
\end{align}
where $N_{i}$ is the number of particle $i$ being emitted, $g_{i}$
denotes its multiplicity, $\gamma_{\text{gray}}$ is the graybody
factor, $T_{{\rm BH}}$ is the temperature of the black hole, $E$
is the particle energy,  and $\eta$ takes $1$ or $-1$ for Bose-Einstein
or Fermi-Dirac statistics, respectively.\footnote{ One may also  take Maxwell-Boltzmann statistics, corresponding to
$\eta=0$, as a useful approximation since  it usually simplifies
analytical calculations significantly. }

The PBH temperature is related to its mass $m_{\text{BH}}$ via
\begin{equation}
T_{{\rm BH}}=\frac{m_{{\rm pl}}^{2}}{8\pi m_{{\rm BH}}}\thinspace,\label{eq:PBH}
\end{equation}
with $m_{{\rm pl}}=1.22\times10^{19}$ GeV the Planck mass. As the
PBH keeps evaporating and losing energy, its mass  decreases as follows~\cite{Baldes:2020nuv}:
\begin{equation}
m_{{\rm BH}}=m_{{\rm BH0}}\left(1-\frac{t-t_{F}}{\tau_{{\rm BH}}}\right)^{1/3},\ \ t\in[t_{F},\ t_{{\rm ev}}]\thinspace.\label{eq:PBH-3}
\end{equation}
Here $m_{{\rm BH0}}$ is the initial mass of the PBH at $t=t_{F}$
with $t_{F}$ the PBH formation time, $\tau_{{\rm BH}}$ is the lifetime
of the PBH, and $t_{{\rm ev}}\equiv t_{F}+\tau_{{\rm BH}}$ is the
evaporation time.   We refer to Appendix~\ref{sec:PBH-basic} for
a brief review of these basic quantities of PBHs. 

From Eqs.~\eqref{eq:PBH-3} and \eqref{eq:PBH}, one can see that
when the PBH approaches the end of its lifetime (i.e.~$t\to t_{{\rm ev}}$,
$m_{{\rm BH}}\to0$), its temperature $T_{{\rm BH}}$ rises significantly,
leading to the emission of highly energetic radiation. However, the
number of particles emitted during this final stage is relatively
small compared to the total number of particles produced throughout
the entire evaporation process. To quantitatively  study this, one
can integrate Eq.~\eqref{eq:PBH-1} over the full lifetime of the
PBH to obtain the cumulative energy spectrum,
\begin{equation}
\frac{dN_{i}}{dE}=\int_{t_{F}}^{t_{\text{ev}}}\frac{d^{2}N_{i}}{dtdE}dt\thinspace.\label{eq:-24}
\end{equation}

For later convenience, we define 
\begin{equation}
x\equiv\frac{E}{T_{{\rm BH0}}}\thinspace,\ \ y\equiv\frac{p}{T_{{\rm BH0}}}\thinspace,\label{eq:-38}
\end{equation}
where $T_{{\rm BH0}}$ is the initial temperature of the PBH. 

Assuming that the particle mass is negligible and using the Boltzmann
approximation, the integral in Eq.~\eqref{eq:-24} can be computed
analytically, yielding the following cumulative energy spectrum~\cite{Schmitz:2023pfy}:
\begin{equation}
\frac{dN_{i}}{dE}\approx\frac{243g_{i}\tau_{{\rm BH}}}{16\pi^{3}}\left[\frac{1}{x^{3}}-\frac{\Gamma(5,x)}{24x^{3}}\right],\label{eq:-23}
\end{equation}
where $\Gamma(5,x)=e^{-x}\left(x^{4}+4x^{3}+12x^{2}+24x+24\right)$
is an {\it incomplete gamma function}.

For $x\ll1$, Eq.~\eqref{eq:-23} is proportional to $x^{2}$. For
$x\gg1$, the $\Gamma(5,x)$ term can be neglected and Eq.~\eqref{eq:-23}
is proportional to $x^{-3}$, implying   a high-energy tail $\propto E^{-3}$,
which is significantly harder than a generic thermal spectrum. 

\section{High-energy radiations in the early universe \label{sec:High-E-in-universe}}

 The spectral shape discussed in Sec.~\ref{sec:The-hard-spectrum}
is only for PBH evaporation in static spacetime without thermal backgrounds.
For high-energy radiations emitted by PBHs in the early universe,
which undergoes Hubble expansion and contains a dense thermal plasma,
 we should also consider the cosmological redshift effect and possible
thermal wash-out of the high-energy tail.  These effects can be taken
into account by solving the following Boltzmann equation
\begin{equation}
\left[\frac{\partial}{\partial t}-Hp\frac{\partial}{\partial p}\right]f_{i}\left(t,\ p\right)\approx\Gamma_{i,\thinspace{\rm prod}}-\Gamma_{i,\thinspace{\rm abs}}f_{i}\thinspace,\label{eq:-25}
\end{equation}
where $f_{i}(t,\ p)$ is the phase space distribution function of
particle $i$, $H=a^{-1}da/dt$ is the Hubble parameter with $a$
the scale factor, $\Gamma_{i,\thinspace{\rm prod}}$ denotes the production
rate of $i$ including thermal processes and PBH evaporation, and
$\Gamma_{i,\thinspace{\rm abs}}$ denotes the absorption rate in the
plasma. 

Although Eq.~\eqref{eq:-25} in principle can be applied to rather
generic circumstances, we are interested in the situation that $f_{i}$
is out of equilibrium due to low thermal reaction rates. For neutrinos,
this implies that only those produced from PBHs after neutrino decoupling
are important to our analyses. We will come back to  the decoupling
issue later in Sec.~\ref{subsec:washout}. Here let us first consider
that the $\Gamma_{i,\thinspace{\rm abs}}$ term in Eq.~\eqref{eq:-25}
is negligible, for which $f_{i}$ can be computed (see Appendix~\ref{sec:Analytical-solutions}
for details) via  
\begin{equation}
f_{i}\left(t,\ p\right)=\int_{0}^{a}\frac{\Gamma_{i,\thinspace{\rm prod}}(a',\ p')}{H(a')a'}da'\thinspace,\label{eq:f-int}
\end{equation}
where $p'\equiv pa/a'$. For PBH evaporation, the corresponding production
rate reads
\begin{equation}
\Gamma_{i,\thinspace{\rm prod}}^{(\text{PBH}\to i)}\approx n_{{\rm BH}}\frac{(2\pi)^{3}}{4\pi p^{2}}\frac{d^{2}N_{i}}{dtdp}\thinspace,\label{eq:-26}
\end{equation}
where $n_{{\rm BH}}$ is the PBH number density in the early universe,
and $\frac{d^{2}N_{i}}{dtdp}=\frac{d^{2}N_{i}}{dtdE}\frac{dE}{dp}$
with $\frac{d^{2}N_{i}}{dtdE}$ given by Eq.~\eqref{eq:PBH-1}. The
factor $\frac{(2\pi)^{3}}{4\pi p^{2}}$ comes from that the number
density is given by $n_{i}=\int f_{i}\frac{4\pi p^{2}}{(2\pi)^{3}}dp$. 

The PBH number density  in the expanding universe simply scales as
\begin{equation}
n_{{\rm BH}}=n_{{\rm BH0}}\frac{a_{F}^{3}}{a^{3}}\thinspace,\ \ a\in[a_{F},\ a_{{\rm ev}}]\thinspace,\label{eq:-27}
\end{equation}
where $n_{{\rm BH0}}$ denotes the initial PBH number density, and
$a_{F}$ and $a_{{\rm ev}}$ are the scale factors at $t=t_{F}$ and
$t=t_{{\rm ev}}$. The initial PBH abundance at formation is typically
parametrized by
\begin{equation}
\beta\equiv\left.\frac{\rho_{{\rm BH0}}}{\rho_{{\rm tot}}}\right|_{t=t_{F}}\thinspace,\label{eq:-28}
\end{equation}
where $\rho_{{\rm BH0}}=n_{{\rm BH0}}\thinspace m_{{\rm BH0}}$ is
the initial PBH energy density and $\rho_{{\rm tot}}$ is the total
energy density of the universe. Throughout this work, we only consider
PBHs with a monochromatic mass function. 

The initial number density $n_{{\rm BH0}}$ can be determined from
$m_{{\rm BH0}}$ and $\beta$ by
\begin{equation}
n_{{\rm BH0}}=\frac{3\beta\gamma^{2}m_{{\rm pl}}^{6}}{32\pi m_{{\rm BH0}}^{3}}\thinspace,\label{eq:-29}
\end{equation}
where $\gamma\approx0.2$ is a factor describing the efficiency of
local overdensities collapsing into PBHs~\cite{Carr:1975qj}.

\subsection{The energy spectrum of a  massless species}

Substituting Eqs.~\eqref{eq:-26} and \eqref{eq:-27} into Eq.~\eqref{eq:f-int},
one can work out the integral analytically  by combining low- and
high-energy limits together---see Appendix~\ref{sec:Analytical-solutions}
for the detailed calculation.  Assuming that the PBH lifespan is
completely contained  in a radiation dominated (RD) era\footnote{This requires the PBH mass $m_{{\rm BH0}}$ to be below $\sim10^{13}$
gram.\label{fn:10-13}}, the result reads
\begin{equation}
\left.f_{i}\right|_{t\to t_{{\rm ev}}}\approx f_{i0}\left[\frac{1-\Gamma(5,x)/24}{x^{5}}+f_{i,\text{IR}}\right],\label{eq:-30}
\end{equation}
with
\begin{equation}
f_{i0}\equiv\frac{3^{6}}{4}\sqrt{3\gamma g_{{\rm BH}}}\beta g_{i}\frac{m_{{\rm pl}}}{m_{{\rm BH0}}}\thinspace,\label{eq:-31}
\end{equation}
and
\begin{equation}
f_{i,\text{IR}}=\frac{1}{36x}\begin{cases}
\log\left(e^{x}+1\right)-x-\kappa_{+}xe^{-x} & (\text{Fermi-Dirac})\\
x-\log\left(e^{x}-1\right)-\kappa_{-}xe^{-x} & (\text{Bose-Einstein})\\
e^{-x}-\kappa_{0}xe^{-x} & (\text{Maxwell-Boltzmann})
\end{cases}\thinspace,\label{eq:-32}
\end{equation}
where $\left(\kappa_{+},\ \kappa_{-},\ \kappa_{0}\right)=\left(\frac{3}{4}-\frac{3\zeta(3)}{16},\ \frac{3}{4}-\frac{\zeta(3)}{4},\ \frac{1}{2}\right)\approx\left(0.52,\ 0.45,\ 0.5\right)$. 

\begin{figure}
\centering

\includegraphics[width=0.98\textwidth]{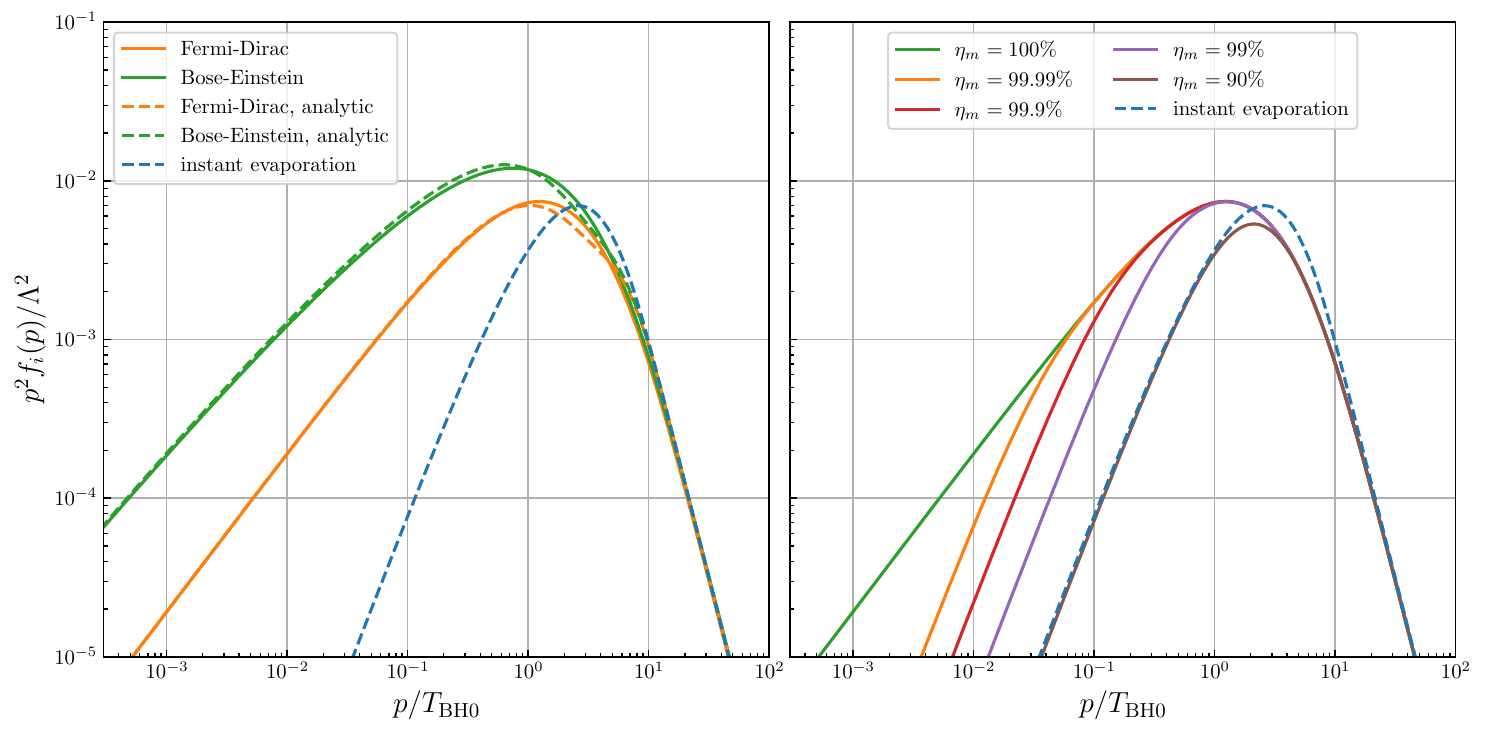}

\caption{The energy spectrum of a massless species emitted from PBH evaporation
in the early universe. Here $\Lambda^{2}\equiv T_{{\rm BH0}}^{2}f_{i0}$
with $f_{i0}$ given by Eq.~\eqref{eq:-31} and $\eta_{m}$ in the
right panel denotes the percentage of the remaining mass of the PBHs
to be included in the calculation---see Eq.~\eqref{eq:-49} for the
definition. \label{fig:f-spectrum}}
\end{figure}

In Eq.~\eqref{eq:-31}, $g_{{\rm BH}}\equiv\frac{m_{{\rm BH0}}^{3}}{3m_{{\rm pl}}^{4}\tau_{{\rm BH}}}$
is around $7.5\times10^{-3}$ if all SM degrees of freedom are taken
into account. The $f_{i,\text{IR}}$ term in Eq.~\eqref{eq:-30} becomes
dominant when $x\ll1$. Hence we refer to it as the infrared (IR)
contribution. This part depends very significantly on the quantum
statistics of the emitted particles\,---\,see the left panel of
Fig.~\ref{fig:f-spectrum} where we plot $p^{2}f_{i}$ according
to Eq.~\eqref{eq:-30}, with $f_{i,\text{IR}}$ taking the Fermi-Dirac
or Bose-Einstein form in Eq.~\eqref{eq:-32} (dashed lines), and compare
it with the full numerical results (solid lines). In contrast to the
IR part, the ultraviolet (UV) part is insensitive to the difference
between Fermi-Dirac and  Bose-Einstein statistics. 

For comparison, we also plot the result for instant evaporation (blue
dashed), assuming that the PBHs evaporate instantly at $t=t_{{\rm ev}}$.
If the PBH could evaporate instantly without a significant life span
in the expanding  universe, the cosmological redshift effect on $f_{i}$
before $t_{{\rm ev}}$ could be neglected, leading to $f_{i}\propto x^{-2}\frac{dN_{i}}{dE}$
with $\frac{dN_{i}}{dE}$ the cumulative spectrum in Eq.~\eqref{eq:-23}.
Therefore, the instant evaporation curve corresponds to the $1-\Gamma(5,x)/24$
term in Eq.~\eqref{eq:-30}. 

In the right panel of Fig.~\ref{fig:f-spectrum}, we show that only
a quite small fraction of the PBH mass at the early phase of evaporation
is responsible for the IR contribution. Here we introduce a parameter
$\eta_{m}$ defined as  
\begin{equation}
\eta_{m}\equiv\frac{m_{{\rm BH}}(t=t_{m})}{m_{{\rm BH0}}}\thinspace,\label{eq:-49}
\end{equation}
where $t_{m}\in[t_{F},\ t_{{\rm ev}}]$ is an intermediate time point.
The $\eta_{m}$ parameter quantifies the fraction of the remaining
mass of the PBHs at the intermediate point. The curves with $\eta_{m}<1$
only include  the contribution of evaporation during $[t_{m},\ t_{{\rm ev}}]$.
For instance, the difference between the red ($\eta_{m}=99.9\%$)
and the green ($\eta_{m}=100\%$) curves  can be interpreted as the
contribution of  the first $0.1\%$ of the PBH mass evaporating during
$[t_{F},\ t_{m}]$. 

From Fig.~\ref{fig:f-spectrum}, we can  draw the conclusion that
the high-energy part ($p/T_{{\rm BH0}}\gg1$) is very insensitive
to the early phase of evaporation  and also independent of the quantum
statistics of the particles emitted. 

The number density  of  particles after evaporation can be computed
by integrating Eq.~\eqref{eq:-30}  over the momentum space:  
\begin{equation}
n_{i}\equiv\int f_{i}\frac{4\pi p^{2}}{(2\pi)^{3}}dp=f_{i0}\frac{c_{\pm}T_{{\rm BH0}}^{3}}{48\pi^{2}}\thinspace,\ \ (t=t_{{\rm ev}})\thinspace,\label{eq:-44}
\end{equation}
where $c_{\pm}=\frac{3\zeta(3)}{4}$ and $\zeta(3)$ for Fermi-Dirac
and Bose-Einstein statistics, respectively. The Boltzmann approximation
corresponds to $c_{\pm}\to1$. Eq.~\eqref{eq:-44} can also be obtained
via $n_{i}=n_{{\rm BH}}N_{i}$ where $N_{i}$ is the total number
of $i$ particles emitted by each PBH---see Eq.~\eqref{eq:Ni} in
Appendix~\ref{sec:PBH-basic}. This approach leads to the same result
as Eq.~\eqref{eq:-44}.

\subsection{The energy spectrum of a stable massive species}

For a massive species emitted by PBHs, if its mass $m_{i}$ is well
below $T_{{\rm BH0}}$, then the emission rate itself is not significantly
affected by the mass. However, the IR part of the energy spectrum
at $t=t_{{\rm ev}}$ can be significantly altered by the mass when
those particles emitted at an early stage are red-shifted to non-relativistic. 

The calculation of $f_{i}$ in this case is straightforward using
Eq~\eqref{eq:f-int}. However, we do not find simple analytical results
for nonzero $m_{i}\lesssim T_{{\rm BH0}}$. Numerically, we find that
the high-energy part of the spectrum is insensitive to the mass effect
while the low-energy part typically varies within one order of magnitude
when $m_{i}/T_{{\rm BH0}}$ varies from $0$ to $1$. 

If the mass is well above $T_{{\rm BH0}}$, then the emission rate
would be suppressed until the increasing $T_{{\rm BH}}$ reaches  $m_{i}$.
We denote the time at this moment by $t_{m_{i}}$.  For simplicity,
we neglect the contribution of the suppressed emission and only take
into account the emission when $T_{{\rm BH}}\gtrsim m_{i}$. This
implies that when computing $f_{i}$ using Eq.~\eqref{eq:f-int},
we actually integrate over the period of $t\in[t_{m_{i}},\ t_{{\rm ev}}]$
with
\begin{equation}
t_{m_{i}}=t_{{\rm ev}}-\tau_{{\rm BH}}\mu^{-3}\thinspace,\ \ \text{for }\mu>1\thinspace,\label{eq:-33}
\end{equation}
where $\mu\equiv m_{i}/T_{{\rm BH0}}=\sqrt{x^{2}-y^{2}}$.   

In this approach, we find that the resulting $f_{i}$ can be very
accurately computed using the instant evaporation approximation, which
leads to
\begin{equation}
\left.f_{i}\right|_{t\to t_{{\rm ev}}}\approx f_{i0}\frac{y}{x^{6}}\left[1-\frac{1}{24}\Gamma\left(5,\frac{x}{\max(\mu,1)}\right)\right].\label{eq:-34}
\end{equation}
Here  $\max(\mu,1)$ arises from the requirement that Eq.~\eqref{eq:-33}
should be used only when $\mu>1$. In the limit of $\mu\to0$, Eq.~\eqref{eq:-34}
recovers the instant evaporation part of Eq.~\eqref{eq:-30}. In the
limit of $x\gg\mu$, the incomplete gamma function can be neglected,
which implies that the high-energy limit of Eq.~\eqref{eq:-34} is
the same as the massless case. This is true even for heavy species
with masses significantly higher than $T_{{\rm BH0}}$. 

The number density of  particles in this case can be estimated using
Eq.~\eqref{eq:-44} if $m_{i}\lesssim T_{{\rm BH0}}$, or by multiplying
it with a correction factor $T_{{\rm BH0}}^{2}/m_{i}^{2}$ if $m_{i}\gtrsim T_{{\rm BH0}}$.

\subsection{The energy spectrum of a slow-decaying species\label{subsec:spectrum-decay}}

\begin{figure}
\centering

\includegraphics[width=0.85\textwidth]{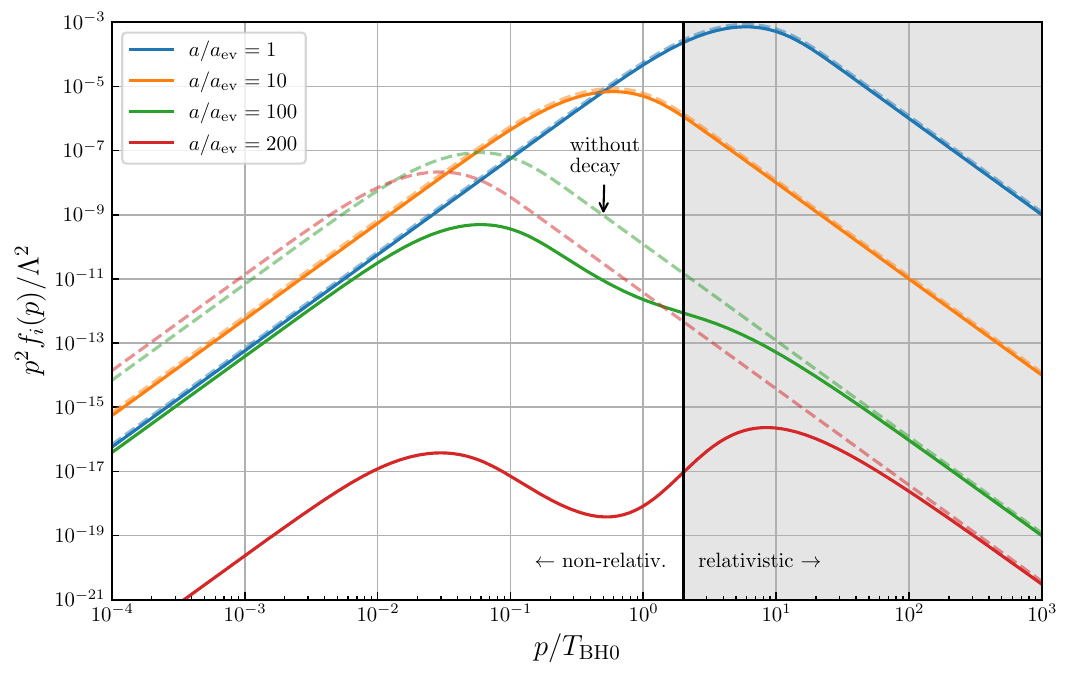}

\caption{The energy spectrum of a slow-decaying species after being produced
from PBHs in the early universe. Here $\Lambda^{2}\equiv T_{{\rm BH0}}^{2}f_{i0}$
with $f_{i0}$ given by Eq.~\eqref{eq:-31}. The shown example assumes
$\mu=2.0$ and $\Gamma_{iD}=10^{-3}H_{{\rm ev}}$ (solid lines) or
$\Gamma_{iD}=0$ (dashed). Note that the differences between solid
and dash lines for $a/a_{{\rm ev}}\lesssim10^{3/2}$ are actually
invisibly small. For better illustration, we have increased the dashed
curves by $20\%$. The gray region corresponds to  $p>\mu$, implying
that the particles in this region are relativistic.  \label{fig:f-spectrum-decay}}
\end{figure}

Next, let us consider a slow-decaying massive species. By ``slow-decaying'',
we mean that its decay rate at rest, denoted by $\Gamma_{iD}$, satisfies
\begin{equation}
\Gamma_{iD}\ll H_{{\rm ev}}\thinspace,\label{eq:-35}
\end{equation}
where $H_{{\rm ev}}\equiv H(t_{{\rm ev}})$.   Based on Eq.~\eqref{eq:-35},
we assume that the PBH produced particles do not decay before $t_{{\rm ev}}$.
Under this assumption, the subsequent evolution of $f_{i}$ after
$t_{{\rm ev}}$ is given by
\begin{equation}
f_{i}(t,p)=f_{i}\left(t_{{\rm ev}},\frac{a_{t}}{a_{{\rm ev}}}p\right)e^{-\Gamma_{iD}\tilde{t}}\thinspace,\ \ \tilde{t}\equiv\int_{t_{\text{ev}}}^{t}\frac{m_{i}{\rm d}\tau}{\sqrt{\left(pa_{t}/a_{\tau}\right)^{2}+m_{i}^{2}}}\thinspace,\label{eq:f-decay}
\end{equation}
where $a_{t}\equiv a(t)$ and $a_{\tau}\equiv a(\tau)$. The derivation
of Eq.~\eqref{eq:f-decay} is presented in Appendix~\ref{sec:decay}.
One can check that Eq.~\eqref{eq:f-decay} is indeed a solution of
the Boltzmann equation by substituting it back into  Eq.~\eqref{eq:-25}
with $\Gamma_{i,\thinspace{\rm prod}}=0$ and
\begin{equation}
\Gamma_{i,\thinspace{\rm abs}}=\frac{m_{i}}{\sqrt{m_{i}^{2}+p^{2}}}\Gamma_{iD}\thinspace.\label{eq:-36}
\end{equation}

In the radiation-dominated era, where $t\propto a^{2}$, the integral
in Eq.~\eqref{eq:f-decay} can be worked out analytically (see Appendix~\ref{sec:decay}
for detailed calculations), leading to
\begin{equation}
\frac{\tilde{t}}{t}=\frac{x}{\mu}-\frac{x_{\text{ev}}}{\mu}\frac{a_{{\rm ev}}^{2}}{a_{t}^{2}}+\frac{y^{2}}{2\mu^{2}}\ln\frac{(x-\mu)(x_{\text{ev}}+\mu)}{(x+\mu)(x_{\text{ev}}-\mu)}\thinspace,\label{eq:-37}
\end{equation}
where $x_{{\rm ev}}\equiv\sqrt{\left(ya_{t}/a_{{\rm ev}}\right)^{2}+\mu^{2}}$. 

In Fig.~\ref{fig:f-spectrum-decay},  we plot an example with $\mu=2$
and $\Gamma_{iD}=10^{-3}H_{{\rm ev}}$.  The curves are obtained
using Eq.~\eqref{eq:f-decay}, where $f_{i}\left(t_{{\rm ev}},p\right)$
is given by Eq.~\eqref{eq:-34} and $\tilde{t}$ is given by Eq.~\eqref{eq:-37}.
For the shown example, since $\Gamma_{iD}/H\approx2t\Gamma_{iD}\approx10^{-3}(a/a_{{\rm ev}})^{2}$,
the blue and orange curves correspond to $2t\Gamma_{iD}\approx10^{-3}$
and $10^{-1}$, respectively. Therefore, the evolution from the blue
curve to the orange is simply a  redshift effect. When $\Gamma_{iD}/H\approx2t\Gamma_{iD}$
exceeds unity, the decay effect become significant, as can be seen
from the difference between the dashed (without decay) and solid (including
decay) curves.  Here one can see that the non-relativistic part decays
much faster than the relativistic part in the gray region. It is interesting
to note that further evolution may cause a double-peak structure due
to the interplay between  redshift and decay. 

Let us comment here that for slow-decaying particles considered in
this work, the thermal production of these particles  is assumed to
be negligible because the small decay rate implies very weak couplings
to SM particles, rendering the thermal production very inefficient.
The production of slow-decaying particles from PBHs, however, is not
limited by the weak couplings and accounts for the abundance of these
particles in the early universe.  The weak couplings also ensure
that  the  characteristic power-law tail in the high-energy part
of the spectrum is not washed out by the thermal bath.

\subsection{The energy spectrum of a decay product \label{subsec:decay-product}}

The decay product of the slow-decaying species considered in Sec.~\ref{subsec:spectrum-decay}
may also have interesting phenomenology. If the slow-decaying species
is heavy and decays non-relativistically, the particles produced from
the decay are much more energetic than those directly produced from
PBHs, because the energy stored in the mass of the heavy species is
resistant to cosmological redshift until it is converted into radiation. 

Taking the example in Fig.~\ref{fig:f-spectrum-decay} with $\Gamma_{iD}=10^{-3}H_{{\rm ev}}$
and $\mu=2.0$, one can see that the heavy species does not decay
immediately after evaporation. After cosmological expansion by a factor
of $a/a_{{\rm ev}}\sim10^{3/2}$, when $H$ becomes comparable to
$\Gamma_{iD}$, the decay effectively starts, producing light particles
with the energy $\sim\mu T_{{\rm BH0}}/2=T_{{\rm BH0}}$. If these
light particles were directly produced from PBH, then their typical
energy after the redshift $a/a_{{\rm ev}}$ should be $T_{{\rm BH0}}a_{{\rm ev}}/a\sim0.03T_{{\rm BH0}}$.
Therefore, the decay product in this example is a factor of $30$
more energetic than those directly produced from PBHs.  

To compute the energy spectrum of particles produced from the decay
of a heavy species emitted by PBH, one can again use the Boltzmann
equation \eqref{eq:-25}, with the right-hand side including decay
as a production term. Considering the process $i\to j+k$ where $j$
and $k$ are two generic light species, the Boltzmann equation for
$j$  can be solved by
\begin{equation}
f_{j}(t,p_{j})=\frac{m_{i}\Gamma_{iD}}{p_{j}^{2}a_{t}^{2}}\int_{t_{{\rm ev}}}^{t}a_{\tau}^{2}{\rm d}\tau\int_{p_{i}^{\min}}^{\infty}{\rm d}p_{i}\frac{p_{i}}{E_{i}}f_{i}(\tau,p_{i})\thinspace,\label{eq:-39}
\end{equation}
where 
\begin{equation}
p_{i}^{\min}\equiv\left|\frac{m_{i}a_{\tau}}{4p_{j}a_{t}}-\frac{p_{j}a_{t}}{a_{\tau}}\right|.\label{eq:-40}
\end{equation}
 Eq.~\eqref{eq:-39} can be derived by combining the calculations
in Appendices~\ref{sec:Analytical-solutions} and \ref{sec:collision}.

In general, $f_{j}$ can be computed by feeding the result of $f_{i}(t,p)$
obtained from Eq.~\eqref{eq:f-decay} into Eq.~\eqref{eq:-39} and
perform the integrations numerically. There are, however, two circumstances
that one can calculate $f_{j}$ analytically: $f_{i}$ being non-relativistic
or ultra-relativistic when the decay takes effect. The latter leads
to an $E^{-3}$ tail of the energy spectrum of $j$ at the same order
of magnitude as that from direct evaporation. The former is phenomenologically
more interesting due to the aforementioned effect that the energy
store in the heavy mass is resistant to cosmological redshift.  So
below we focus on the non-relativistic case.

When $f_{i}$ decays non-relativistically, we find
\begin{equation}
f_{j}(t,p_{j})\approx16\pi^{2}n_{i{\rm ev}}a_{{\rm ev}}^{3}\left(\frac{2p_{j}a_{t}}{m_{i}a_{d}}\right)^{r}\left(\frac{1}{2p_{j}a_{t}}\right)^{3}\exp\left[-\frac{1}{r}\left(\frac{2p_{j}a_{t}}{m_{i}a_{d}}\right)^{r}\right],\label{eq:-41}
\end{equation}
with 
\begin{equation}
r=\begin{cases}
2 & (a_{d}\in\text{RD})\\
3/2 & (a_{d}\in\text{MD})
\end{cases}\thinspace.\label{eq:-42}
\end{equation}
Here $a_{d}$ is the scale factor at $H=\Gamma_{iD}$, and $n_{i{\rm ev}}$
is the number density of $i$ at $a=a_{{\rm ev}}$.   The index
$r$ depends on whether $a_{d}$ is in the radiation-dominated (RD)
or matter-dominated (MD) era. Note that in practice, one may encounter
the situation that $a_{{\rm ev}}$, $a_{d}$, and $a_{t}$ in Eq.~\eqref{eq:-41}
may be in different eras. In this case,  Eq.~\eqref{eq:-41} is still
valid, because (i) the comoving number density $n_{i}a^{3}$ is almost
constant before the decay takes effect, and (ii) after decay, the
subsequent evolution of the distribution is simply to rescale the
momentum $p_{j}$, which has been accounted for by the combination
$p_{j}a_{t}$, independent of which era $a_{t}$ is in. Indeed, $p_{j}$
appearing in Eq.~\eqref{eq:-41} is always accompanied with $a_{t}$.

Integrating Eq.~\eqref{eq:-41} with respect to $p_{j}$, we obtain
the number density of $j$: 
\begin{equation}
n_{j}=\int_{0}^{m_{i}/2}f_{j}(t,p)\frac{4\pi p^{2}}{(2\pi)^{3}}dp=\frac{1}{a^{3}}n_{i{\rm ev}}a_{{\rm ev}}^{3}\left(1-e^{-\Gamma_{iD}t}\right),\label{eq:-43}
\end{equation}
which implies that the total comoving number density $\left(n_{i}+n_{j}\right)a^{3}$
is conserved. 

\subsection{Condition for free-streaming\label{subsec:washout}}

So far, our calculations have been derived under the assumption that
these particles after being produced from the PBHs   stream freely
through the thermal plasma of the early universe. For neutrinos, the
free-streaming assumption is valid only after the universe has cooled
down to the neutrino decoupling temperature, $T_{\nu{\rm dec}}\approx2\ \text{MeV}$.
 Well before $2$ MeV, neutrinos are in thermal equilibrium so any
deviations from their thermal spectrum would be quickly washed out
by frequent collisions of neutrinos with themselves or other particles
in the thermal bath. After 2 MeV, it is generally assumed that most
neutrinos start free-streaming without interacting with any particles
on their paths. However, very energetic neutrinos may still have sufficiently
high interaction rates which  inhibit them from free streaming. 

To estimate the condition for free streaming, one can compare the
interaction rate with the Hubble expansion rate $H$. The interaction
rate of a high-energy neutrino with the thermal neutrino background
is roughly given by 
\begin{equation}
\Gamma_{\nu}\sim G_{F}^{2}T_{\nu}^{4}E_{\nu}\thinspace,\label{eq:-45}
\end{equation}
where $G_{F}$ is the Fermi constant, $E_{\nu}$ is the energy of
the high-energy neutrino, and $T_{\nu}$ is the temperature of thermal
neutrinos. Comparing $\Gamma_{\nu}$ with $H\approx6T_{\nu}^{2}/m_{{\rm pl}}$,\footnote{The exact value of $Hm_{{\rm pl}}/T_{\nu}^{2}$ varies from $5.44$
to $5.97$ when the universe evolves from a few MeV to a few eV. Since
here it is only used for an approximate estimate, we take the prefactor
$6$ for simplicity. Note that here $H$ is expressed in terms of
$T_{\nu}$ rather than the photon temperature, $T_{\gamma}$. If the
latter is used, the prefactor would vary from $5.44$ to $3.04$.} which is approximately valid for $T_{\nu}$ within MeV to eV scales,
we obtain 
\begin{equation}
\Gamma_{\nu}\lesssim H\ \ \Leftrightarrow\ \ E_{\nu}\lesssim4\ \text{TeV}\cdot\left(\frac{\text{keV}}{T_{\nu}}\right)^{2}.\label{eq:-46}
\end{equation}
Eq.~\eqref{eq:-46} implies that  when the universe cools down to
keV, high-energy neutrinos with $E_{\nu}$ above $4$ TeV still cannot
free-stream. Therefore, if PBHs produce high-energy neutrinos with
the aforementioned $E_{\nu}^{-3}$ spectrum in the keV universe, the
tail above 4 TeV would be washed out. We refer to the upper bound
on $E_{\nu}$ in Eq.~\eqref{eq:-46} as the wash-out point, above
which the high-energy spectrum should be cut off. Note that this wash-out
point undergoes cosmological redshift in the subsequent evolution.
The present-day value of the wash-out point is approximately given
by 
\begin{equation}
E_{\nu}^{(\text{ws})}\simeq4\ \text{TeV}\cdot\left(\frac{\text{keV}}{T_{\nu}(t_{{\rm ev}})}\right)^{2}\cdot\frac{a_{{\rm ev}}}{a_{{\rm today}}}\thinspace,\label{eq:-46-1}
\end{equation}
where $T_{\nu}(t_{{\rm ev}})$ denotes the neutrino temperature at
evaporation and $a_{{\rm today}}$ is the scale factor today. 

For electromagnetically-interacting particles such as photons and
charged leptons, the wash-out points are much lower. Their high-energy
spectra produced by PBHs in the early universe would be completely
washed out if the PBHs are lighter than $\sim10^{13}$ g (Note that
PBHs lighter than this mass evaporate earlier than {\it recombination}).
    Unlike neutrinos, these electromagnetically-interacting
particles cannot maintain high-energy spectra. Once produced from
PBHs, they quickly lose energies due to interactions with the thermal
plasma and generate energy injection into the latter. This could affect
BBN predictions and cause CMB spectral distortions, which have been
be used to set constraints on PBHs---see e.g.~Fig.~4 of Ref.~\cite{Carr:2020gox}.

For dark particles like dark matter or dark radiation, the wash-out
points can also be estimated in a similar way if their interactions
with the SM thermal bath are given.

\section{High-energy neutrinos from PBHs\label{sec:High-energy-neutrinos}}

\begin{figure}
\centering

\includegraphics[width=0.9\textwidth]{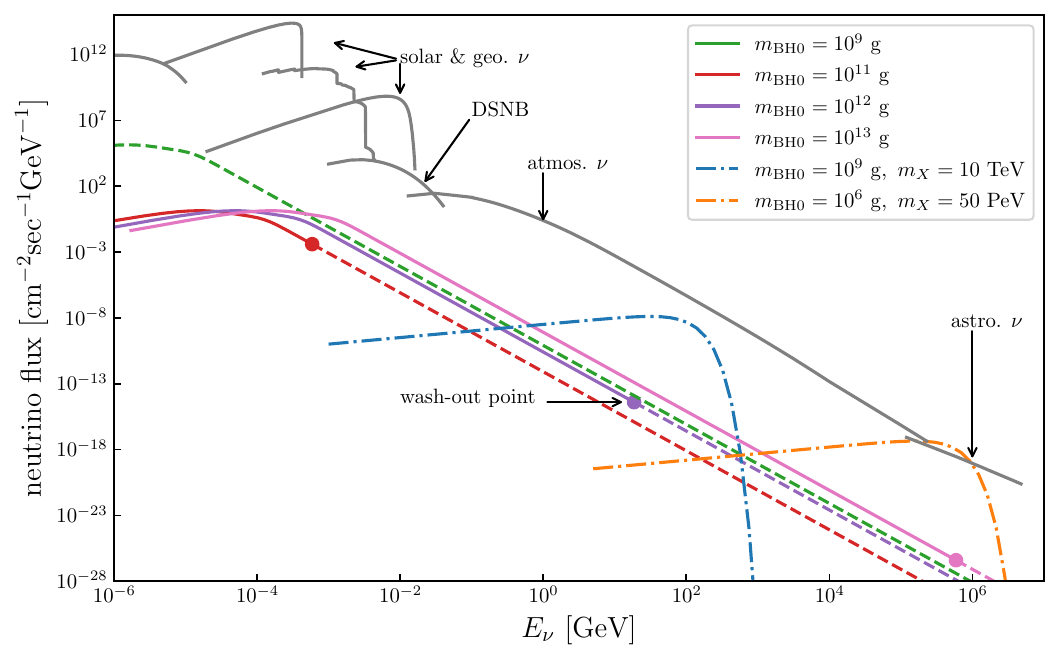}

\caption{Neutrino fluxes produced directly (solid or dashed) or indirectly
(dot-dashed) by PBH evaporation. For direct production, the high-energy
tail beyond the wash-out point given in Eq.~\eqref{eq:-46-1} would
be washed out by thermal processes. Hence  the results are represented
by  joined solid and dashed curves, except for the green curve which
has a wash-out point out of the plot range. For indirect production,
we assume that a new heavy species (denoted by $X$) that barely interacts
with the SM is produced from PBHs and later decays dominantly to neutrinos.
In this case, there is no wash-out effect if the interaction is sufficiently
weak. \label{fig:flux-wide}}
\end{figure}

 In this section, we apply the calculations in Sec.~\ref{sec:High-E-in-universe}
to neutrinos, focusing primarily on the observability of high-energy
neutrino fluxes generated by PBHs in the present universe.  

Let us first consider the neutrino flux directly produced from PBHs
 evaporating in the early universe. Since neutrinos are almost massless
particles, we adopt Eq.~\eqref{eq:-30} to compute the phase space
distribution $f_{\nu}(p)$ of the neutrinos emitted by the PBHs. Note
that Eq.~\eqref{eq:-30} only gives the distribution at evaporation,
while the present distribution  should be red-shifted to 
\begin{equation}
\left.f_{\nu}(p)\right|_{t\to t_{{\rm today}}}=\left.f_{\nu}\left(p\frac{a_{{\rm today}}}{a_{{\rm ev}}}\right)\right|_{t\to t_{{\rm ev}}}\thinspace.\label{eq:-47}
\end{equation}
After $f_{\nu}$ is obtained, the differential neutrino flux can be
computed via\footnote{Here $E_{\nu}^{2}/(2\pi^{2})=4\pi p^{2}/(2\pi)^{3}$ implies that
the flux is counted bi-directionally, i.e.~neutrinos hitting on a
surface of unit area from both sides are counted. }
\begin{equation}
\frac{d\Phi_{\nu}}{dE_{\nu}}=\frac{E_{\nu}^{2}}{2\pi^{2}}f_{\nu}\thinspace,\label{eq:-48}
\end{equation}
where $\Phi_{\nu}$ is defined as the number of neutrinos streaming
through a detector per unit time per unit area, and the neutrino energy
$E_{\nu}$ is identical to $p$ in Eq.~\eqref{eq:-47}. 

There are two free parameters in the calculation, namely the PBH mass
$m_{\text{BH0}}$ and the abundance parameter $\beta$. Depending
on the magnitude of $m_{\text{BH0}}$, there are various known bounds
on $\beta$, as summarized in Ref.~\cite{Carr:2020gox}. For PBHs
heavier than $10^{9}$ gram, the most restrictive bound on $\beta$
comes from Big bang nucleosynthesis (BBN)\footnote{See e.g.~Fig.~18 in Ref.~\cite{Carr:2020gox} and Fig.~3 in Ref.~\cite{Keith:2020jww}.}.
For $10^{9}\lesssim m_{{\rm BH0}}/{\rm gram}\lesssim10^{10}$, $\beta$
needs to be below $10^{-18}\sim10^{-19}$ to avoid significant modifications
of the $^{4}\text{He}$ abundance~\cite{Carr:2020gox}. For $10^{10}\lesssim m_{{\rm BH0}}/{\rm gram}\lesssim10^{13}$,
the deuterium-to-hydrogen ratio offers a more constraining bound,
$\beta\lesssim10^{-23}\sim10^{-24}$~\cite{Carr:2020gox}. In Fig.~\ref{fig:flux-wide},
we present four samples with $m_{{\rm BH0}}\in\{10^{9},\ 10^{11},\ 10^{12},\ 10^{13}\}$
gram. The value of $\beta$ is set at $10^{-19}$ for $m_{{\rm BH0}}=10^{9}$
gram and $10^{-24}$ for the remaining, such that the BBN bounds are
satisfied. For comparison, we also show the fluxes of known neutrino
sources in nature, including the Sun, and the Earth (labeled ``solar
\& geo.~$\nu$'' in Fig.~\ref{fig:flux-wide}), the diffuse supernova
neutrino background (DSNB), the atmosphere, and astrophysical sources~\cite{Vitagliano:2019yzm}.

According to the discussion in Sec.~\ref{subsec:washout}, there
are wash-out points for high-energy neutrino fluxes produced in the
early universe. Applying Eq.~\eqref{eq:-46-1} to the four samples,
we obtain $E_{\nu}^{(\text{ws})}\approx6\times10^{-4}$ eV, $0.6$,
$18$ GeV, and $0.6$ PeV. Except for the first sample which has $E_{\nu}^{(\text{ws})}$
well below the plotted energy range, these wash-out points are marked
by round dots in Fig.~\ref{fig:flux-wide}. Beyond the wash-out points,
we plot the fluxes in dashed lines to indicate that these parts would
be washed out by thermal processes. It is worth mentioning here, albeit
beyond the scope of this work, that in the presence of sterile neutrinos
that mix with the SM neutrinos, the wash-out point could be altered
or evaded. 

Next, we consider a new physics scenario that not only evades the
wash-out effect but also significantly enhances the fluxes. As has
been discussed in Sec.~\ref{subsec:decay-product}, if among all
possible particles being emitted by PBH evaporation there are heavy,
dark, and slow-decaying particles, neutrinos produced from the decay
of such particles can be more energetic than those being directly
produced from PBHs. Note that a substantial number of secondary neutrinos
can also be generated via decays of other SM particles emitted from
PBHs. These secondary neutrinos, with relatively low energies, mainly
affect the low-energy part of the neutrino flux while their influence
on the high-energy part can typically be accounted for by a factor
of two \cite{Carr:2009jm,Carr:2020gox}. Here we do not include this
contribution and refer to Refs.~\cite{MacGibbon:1991vc,Halzen:1995hu,Carr:2009jm,Carr:2020gox,Calabrese:2021zfq,Capanema:2021hnm,Bernal:2022swt,Liu:2023cqs,DeRomeri:2024zqs}
for more details.  For concreteness, we denote the heavy particles
by $X$ with mass $m_{X}$, and assume that they decay dominantly
to neutrinos at an epoch much latter than the PBH evaporation. Using
Eq.~\eqref{eq:-41}, we can readily compute the corresponding neutrino
fluxes in this scenario. In Fig.~\ref{fig:flux-wide}, we present
two samples to show the neutrino fluxes obtained from this indirect
production process (dot-dashed lines). For these two samples, we set
$(m_{{\rm BH0}}/{\rm gram},\ \beta,\ m_{X}/\text{TeV})$ at ($10^{9}$,
$10^{-25}$, $10$) and ($10^{6}$, $10^{-27}$, $5\times10^{4}$).
The decay rates are determined by setting  $a_{d}/a_{{\rm today}}=10^{-2}$,
which implies that most of the $X$ particles decay in the matter-dominated
era.  Since such slow-decaying dark particles behave like dark matter
at matter-radiation equality, we need to ensure that they do not make
an excessive contribution to the amount of matter at this epoch. As
we have checked, the above $\beta$ values only lead to a negligibly
small contribution to matter, and meanwhile generate sufficiently
high neutrino fluxes of phenomenological interest.

We note here that the dominant decay channel $X\to\nu\overline{\nu}$
could be accompanied by other channels such as $X\to\nu W^{\pm}e^{\mp}$
and $X\to e^{\pm}e^{\mp}$ generated by electroweak radiative corrections.
These decays would cause a certain amount of electromagnetic energy
injection, which  has been stringently constrained by CMB observations.
We have checked that for the benchmarks considered in this work, the
electromagnetic injection of $X$ does not exceed the CMB bound. More
specifically, given the CMB bound on the electromagnetic decay rate
of dark matter, $\Gamma_{\text{DM}\to{\rm EM}}\lesssim1/\left(10^{25}\ {\rm sec}\right)$
\cite{Poulin:2016anj}, we recast it into the bound on the energy
injection rate at recombination, $d\rho_{{\rm EM}}/dt\lesssim10^{-27}\text{eV}^{4}/{\rm sec}$.
For very heavy $X$,  the branching ratio of its electromagnetic
decay is roughly given by ${\rm Br}_{X\to\text{EM}}\sim\frac{\alpha_{W}}{2\pi}\sim10^{-3}$
where $\alpha_{W}\approx0.016$ is the electroweak analog to the fine-structure
constant in QED.\footnote{Here ${\rm Br}_{X\to\text{EM}}\sim\frac{\alpha_{W}}{2\pi}$ is only
for the bremsstrahlung process $X\to\nu W^{\pm}e^{\mp}$. For the
loop-level process $X\to e^{\pm}e^{\mp}$, its decay rate is smaller
and  model-dependent---see e.g.~\cite{Xu:2020qek,Chauhan:2020mgv,Xu:2023xva}.
Depending on the UV completion, some cancellations in the loop calculation
can render $X\to e^{\pm}e^{\mp}$ very suppressed.} Then the contribution of $X$ to $d\rho_{{\rm EM}}/dt$ is about
$10^{-3}\rho_{X}\Gamma_{X\to\nu\overline{\nu}}$. By requiring that
this is below $10^{-27}\text{eV}^{4}/{\rm sec}$, we obtain $\beta\lesssim10^{-24}$
and $\beta\lesssim10^{-21}$ for the above two benchmarks with $m_{X}=10$
TeV and $50$ PeV, respectively.

\begin{figure}
\centering

\includegraphics[width=0.9\textwidth]{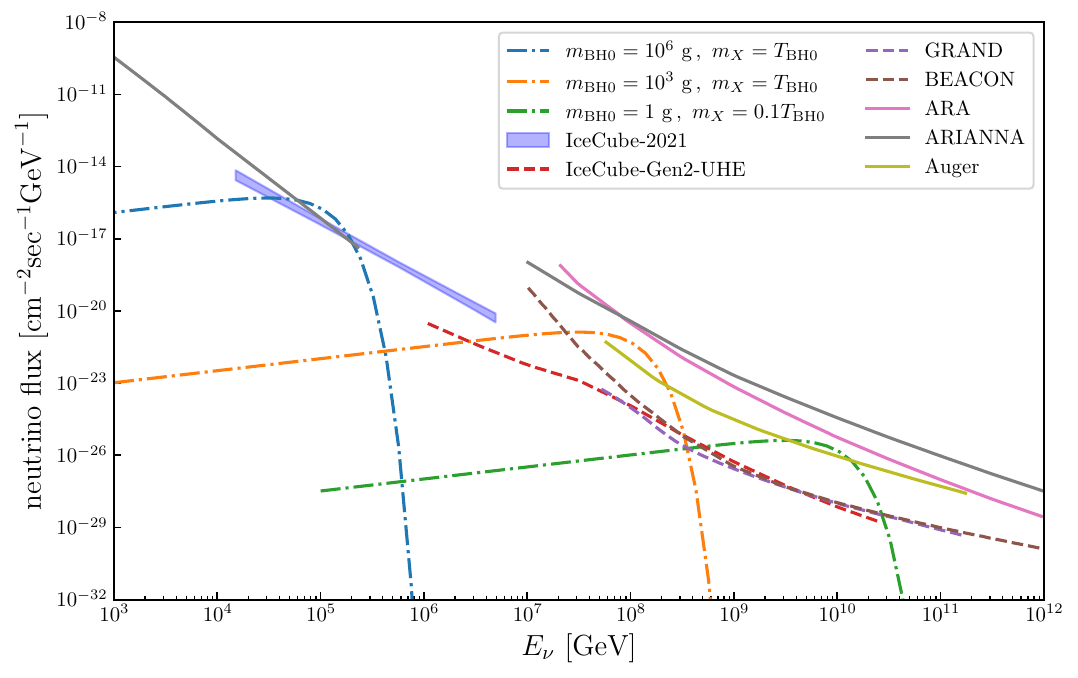}

\caption{Similar to Fig.~\ref{fig:flux-wide} but for the ultra-high-energy
regime. The light blue band labeled IceCube-2021 represents the measured
astrophysical neutrino flux using muon track events only by the IceCube
collaboration~\cite{IceCube:2021uhz}.  The solid curves labeled
ARA, ARIANNA, and Auger are constraints set by these experiments~\cite{ARA:2019wcf,Anker:2019rzo,PierreAuger:2019ens}.
 Dashed lines represent the sensitivity of future experiments~\cite{Ackermann:2022rqc}.
The initial PBH abundance in this figure is set at $\beta=10^{-29}\cdot\left(m_{{\rm BH0}}/\text{g}\right)^{1/3}$.
\label{fig:flux-ultra}}
\end{figure}

In Fig.~\ref{fig:flux-ultra}, we move toward a much more energetic
regime, the so-called ultra-high-energy regime, which covers the EeV
($10^{9}$ GeV) scale or even higher. At such high energies, neutrinos
in our framework can only be produced via the indirect mechanism.
Here we show that gram to kilogram scale PBHs, assisted by the $X$
particle with  $m_{X}$ comparable to $T_{{\rm BH0}}$,  can generically
produce ultra-high-energy neutrinos around or above the EeV scale. 

There have already been very successful observations of high-energy
astrophysical neutrinos by the IceCube experiment. In Fig.~\ref{fig:flux-ultra},
the light blue band labeled IceCube-2021 represents the measured astrophysical
neutrino flux reported by the IceCube collaboration in 2021~\cite{IceCube:2021uhz}.
It spans from $\sim10$ TeV up to a few PeV.  At higher energies,
ARA~\cite{ARA:2019wcf}, ARIANNA~\cite{Anker:2019rzo}, and Auger~\cite{PierreAuger:2019ens}
have set important constraints on the flux, represented by those  solid
lines.  In the near future, a multitude of experiments including
IceCube-Gen2~\cite{IceCube-Gen2:2020qha}, KM3Net~\cite{KM3Net:2016zxf},
GRAND~\cite{GRAND:2018iaj}, BEACON~\cite{Wissel:2020sec}, RET-N~\cite{Prohira:2019glh},
TAMBO~\cite{Romero-Wolf:2020pzh}, and Trinity~\cite{Otte:2019aaf},
will be capable of probing the ultra-high-energy regime with unprecedented
sensitivities\,---\,see Ref.~\cite{Ackermann:2022rqc} for a
recent review. In Fig.~\ref{fig:flux-ultra},  we selectively present
some of the future sensitivity curves.   This figure demonstrates
 that PBHs may serve as   novel ultra-high-energy neutrino sources
of great importance to  future experiments.

\section{Conclusion \label{sec:Conclusion}}

PBHs offer a novel and interesting venue for generating  extremely
energetic radiations.  In this work, we investigate the possibility
of PBHs generating high-energy (TeV\textendash{}PeV) and ultra-high-energy
(PeV\textendash{}EeV) neutrinos in the universe. We focus on PBHs
that can fully evaporate in the early universe and consider two scenarios:
direct and indirect production of neutrinos from PBHs. 

For direct production, we find that the produced neutrinos  feature
a hard energy spectrum $\propto E^{-3}$ in the high-energy limit.
Due to interactions with the thermal bath, the spectrum has a wash-out
point, above which the high-energy neutrinos would lose energy rapidly
in the thermal plasma of the early universe. Taking both the hard
spectrum and the wash-out point into account, we show that PBHs with
a monochromatic mass of $10^{13}$ gram are capable of generating
high-energy neutrinos close to the PeV scale, as is shown in Fig.~\ref{fig:flux-wide}.
The flux obtained in this direct production scenario, however, is
too low for realistic detection. 

The indirect production scenario involves a heavy dark particle $X$,
which is emitted by PBHs and later decays into neutrinos.  We find
that this allows gram to kilogram scale PBHs to generate ultra-high-energy
neutrino fluxes around or above the EeV scale relevant to next-generation
neutrino telescopes such as IceCube-Gen2 and GRAND\,---\,see Fig.~\ref{fig:flux-ultra}.

Our work demonstrates that PBHs evaporating in the early universe
may be responsible for the most energetic radiations observed today.
It also sets an important goal for future experiments searching for
ultra-high-energy neutrinos. 
\begin{acknowledgments}
We thank Yuber F.~Perez-Gonzalez for helpful discussions.  This
work is supported in part by the National Natural Science Foundation
of China under grant No.~12141501 and also by the CAS Project for
Young Scientists in Basic Research (YSBR-099). 
\end{acknowledgments}

\appendix

\section{Basic formulae of PBH evaporation\label{sec:PBH-basic}}

In this appendix,  we briefly review a few useful formulae of PBH
evaporation. 

The emission rate of Hawking radiation has been given by Eq.~\eqref{eq:PBH-1},
which contains the graybody factor $\gamma_{\text{gray}}$. At $E\gg T_{{\rm BH}}$,
the graybody factor approaches the geometrical-optics limit,
\begin{equation}
\gamma_{\text{gray}}\approx\frac{27E^{2}m_{{\rm BH}}^{2}}{m_{{\rm pl}}^{4}}\thinspace.\label{eq:PBH-2}
\end{equation}
At lower energies, the graybody factor exhibits spin-dependent oscillatory
behavior and is  suppressed at $E\ll T_{{\rm BH}}$\,---\,see
Fig.~1 of Ref.~\cite{Cheek:2021odj} for explicit examples. 

From Eq.~\eqref{eq:PBH-1}, one can compute the energy loss rate $\int E\frac{d^{2}N_{i}}{dtdE}dE$
caused by emitting a single species. Adding up the contributions of
all species that can be effectively emitted, one obtains the mass
loss rate of the PBH:
\begin{equation}
\frac{dm_{{\rm BH}}}{dt}=-g_{{\rm BH}}\frac{m_{{\rm pl}}^{4}}{m_{{\rm BH}}^{2}}\thinspace,\label{eq:-20}
\end{equation}
where the dimensionless coefficient $g_{{\rm BH}}$ in the geometrical-optics
limit has been computed in Refs.~\cite{Baldes:2020nuv,Cheek:2021odj}:
\begin{equation}
g_{{\rm BH}}\approx\frac{27}{4}\frac{g_{\star}}{30720\pi}\thinspace,\label{eq:-21}
\end{equation}
with $g_{\star}$ the effective number of degrees of freedom at the
temperature $T_{{\rm BH}}$. 

Under the assumption that $g_{{\rm BH}}$ is a constant during evaporation,
Eq.~\eqref{eq:-20} can be solved analytically:
\begin{equation}
m_{{\rm BH}}=m_{{\rm BH0}}\left(1-\frac{t-t_{F}}{\tau_{{\rm BH}}}\right)^{1/3},\ \ \text{with}\ \ \tau_{{\rm BH}}=\frac{m_{{\rm BH0}}^{3}}{3g_{{\rm BH}}m_{{\rm pl}}^{4}}\thinspace.\label{eq:-22}
\end{equation}
Here $m_{{\rm BH0}}$ is the initial mass of the PBH at $t=t_{F}$
with $t_{F}$ the PBH formation time, and $\tau_{{\rm BH}}$ is the
lifetime of the PBH. 

By integrating Eq.~\eqref{eq:PBH-1} over $t$ and $E$, we obtain
the total number of particles:  
\begin{equation}
N_{i}=\int\frac{d^{2}N_{i}}{dtdE}dEdt=\frac{81g_{i}\tau_{{\rm BH}}T_{{\rm BH0}}}{128\pi^{3}}\cdot\frac{\text{Li}_{3}(\eta)}{\eta}\thinspace,\label{eq:Ni}
\end{equation}
where $\text{Li}_{3}(\eta)$ is the polylogarithm function, with $\eta=-1$,
$1$, and $0$ for Fermi-Dirac, Bose-Einstein, and Maxwell-Boltzmann
statistics, respectively.  It is useful to note that
\begin{equation}
\frac{\text{Li}_{3}(\eta)}{\eta}=\begin{cases}
\frac{3\zeta(3)}{4} & (\text{Fermi-Dirac})\\
\zeta(3) & (\text{Bose-Einstein})\\
1 & (\text{Maxwell-Boltzmann})
\end{cases}\thinspace.\label{eq:-55}
\end{equation}

The formation time $t_{F}$ is given by \cite{Baldes:2020nuv}
\begin{equation}
t_{F}\approx\frac{m_{{\rm BH0}}}{\gamma m_{{\rm pl}}^{2}}\thinspace,\label{eq:-50}
\end{equation}
where $\gamma\approx0.2$ is a factor describing the efficiency of
local overdensities collapsing into PBHs~\cite{Carr:1975qj}. Note
that Eq.~\eqref{eq:-50} is only a rough approximation commonly used
to estimate the formation time. In this work, the value of $t_{F}$
is not important and has very little impact on the results---see
discussions below Eq.~\eqref{eq:-51}. 

In the radiation-dominated universe where $H\approx1/(2t)$, Eq.~\eqref{eq:-50}
implies
\begin{equation}
H(t_{F})=\frac{\gamma m_{{\rm pl}}^{2}}{2m_{{\rm BH0}}}\thinspace,\ \ensuremath{\rho}_{{\rm tot}}(t_{F})=\frac{3\gamma^{2}m_{{\rm pl}}^{6}}{32\pi m_{{\rm BH0}}^{2}}\thinspace,\ T(t_{F})\approx\left(\frac{45\gamma^{2}m_{{\rm pl}}^{6}}{16\pi^{3}g_{\star}m_{{\rm BH0}}^{2}}\right)^{1/4},\ \ (\text{RD})\thinspace.\label{eq:-52}
\end{equation}
Using specific values, the temperature of the universe at $t_{F}$
reads
\begin{equation}
T(t_{F})\approx1.4\times10^{15}\ \text{GeV}\cdot\left(\frac{10\ {\rm g}}{m_{{\rm BH0}}}\right)^{1/2},\ \ (\text{RD})\thinspace.\label{eq:-53}
\end{equation}

From Eqs.~\eqref{eq:-50} and \eqref{eq:-22}, we get
\begin{equation}
\frac{t_{F}}{\tau_{{\rm BH}}}=\frac{3g_{{\rm BH}}m_{{\rm pl}}^{2}}{\gamma m_{\text{BH0}}^{2}}\thinspace,\label{eq:-51}
\end{equation}
which implies $\tau_{{\rm BH}}\gg t_{F}$ for the PBH mass range considered
in this work. Therefore, the PBH evaporation time  $t_{{\rm ev}}\equiv t_{F}+\tau_{{\rm BH}}$
is approximately determined by $\tau_{{\rm BH}}$.  Assuming $t_{{\rm ev}}\approx\tau_{{\rm BH}}$
and using $a\propto t^{1/2}$ in the RD universe, we obtain
\begin{equation}
\left(\frac{a_{F}}{a_{\text{ev}}}\right)\approx\left(\frac{3g_{{\rm BH}}}{\gamma}\right)^{1/2}\frac{m_{{\rm pl}}}{m_{\text{BH0}}}\thinspace,\ \ (\text{RD})\thinspace.\label{eq:-54}
\end{equation}

\section{Integration of $\Gamma_{{\rm prod}}$ \label{sec:Analytical-solutions}}

In the expanding universe, the phase space distribution of a generic
species, denoted by $f\left(t,\ \mathbf{p}\right)$, is governed by
the Boltzmann equation
\begin{equation}
\left[\frac{\partial}{\partial t}-H\mathbf{p}.\nabla_{\mathbf{p}}\right]f\left(t,\ \mathbf{p}\right)=C^{(f)}\thinspace,\label{eq:}
\end{equation}
where $C^{(f)}$ is the collision term which can be written as 
\begin{equation}
C^{(f)}=(1-f)\Gamma_{{\rm prod}}-f\Gamma_{{\rm abs}}\thinspace,\label{eq:-16}
\end{equation}
with $\Gamma_{{\rm abs}}$ and $\Gamma_{{\rm prod}}$ the absorption
and production rates of the species under consideration. 

In this work, we are only concerned with isotropic distributions for
which we can use $f\left(t,\ \mathbf{p}\right)=f\left(t,\ p\right)$
and $H\mathbf{p}.\nabla_{\mathbf{p}}=Hp\partial_{p}$. 

In the limit of $f\ll1$, which implies $C^{(f)}\approx\Gamma_{{\rm prod}}$,
one can obtain the following analytical solution of Eq.~\eqref{eq:}:
\begin{equation}
f\left(t,\ p\right)=\int_{0}^{a}\frac{\Gamma_{{\rm prod}}(a',\ p')}{H(a')a'}da'\thinspace,\label{eq:-17}
\end{equation}
where $p'\equiv pa/a'$. Eq.~\eqref{eq:-17} can be derived by noticing
that the comoving momentum $pa$ is unchanged by the Hubble expansion---see
e.g.~Appendix B of Ref.~\cite{Li:2022bpp} for details.  

For $\Gamma_{{\rm prod}}$ with relatively simple analytic forms,
one can integrate Eq.~\eqref{eq:-17} analytically. Below we apply
Eq.~\eqref{eq:-17} to a few cases with specific forms of $\Gamma_{{\rm prod}}(a,\ p)$. 

\subsection{PBH evaporation with cosmological redshift  }

For massless particles emitted from PBHs in the expanding universe,
the production term is given in Eq.~\eqref{eq:-26}. In the RD era,
using $H(a)=H_{\text{ev}}a_{{\rm ev}}^{2}/a^{2}$, we obtain the following
integrand for Eq.~\eqref{eq:-17}:
\begin{equation}
\frac{\Gamma_{{\rm prod}}(a',\ p')}{H(a')a'}=C_{0}\frac{a_{\text{ev}}\left(1-a'{}^{2}/a_{{\rm ev}}^{2}\right)^{2/3}}{36a'{}^{2}\left(\exp\left[\frac{p'}{T_{{\rm BH0}}}\left(1-a'{}^{2}/a_{{\rm ev}}^{2}\right)^{1/3}\right]-\eta\right)}\thinspace,\label{eq:-56}
\end{equation}
where
\begin{equation}
C_{0}\equiv3^{6}\pi\beta\gamma^{2}g_{i}T_{{\rm BH0}}H_{\text{ev}}^{-1}\left(\frac{a_{F}}{a_{\text{ev}}}\right)^{3}.\label{eq:-57}
\end{equation}
Then Eq.~\eqref{eq:-17} can be written as
\begin{equation}
f=C_{0}\int_{a_{F}/a_{{\rm ev}}}^{1}\frac{R^{2}}{36}\frac{\text{d}r_{a}}{e^{Rx}-\eta}\thinspace,\ \ {\rm with}\ \ R\equiv\text{\ensuremath{\frac{\left(1-r_{a}^{2}\right)^{1/3}}{r_{a}}}}\thinspace,\label{eq:-59}
\end{equation}
where $r_{a}\equiv a'/a_{{\rm ev}}$. 

Since the IR part of $f$ is mainly produced by the first $\sim1\%$
of PBH mass evaporation (see the right panel of Fig.~\ref{fig:f-spectrum})
and the $(1-r_{a}^{2})^{1/3}$ part in $R$ arises from the variation
of $m_{{\rm BH}}$, we can assume that $m_{{\rm BH}}$ is a constant
when producing the IR part. Under this assumption, we can replace
$R\to1/r_{a}$ and compute the integral in Eq.~\eqref{eq:-59} analytically:
\begin{equation}
\left.f\right|_{Rr_{a}\to1}\approx\frac{C_{0}}{36x}\begin{cases}
\log\left(e^{x}+1\right)-x & (\eta=-1)\\
x-\log\left(e^{x}-1\right) & (\eta=1)\\
e^{-x} & (\eta=0)
\end{cases}.\label{eq:-60}
\end{equation}
Eq.~\eqref{eq:-60} is valid only for $x\ll1$. 

On the other hand, since the remaining $\sim90\%$ of $m_{{\rm BH0}}$
is released within a relatively narrow time window close to $t_{{\rm ev}}$,
the instant evaporation result in Eq.~\eqref{eq:-23} should be able
to accurately describe the non-IR part. So in principle, one can add
$\left.f\right|_{Rr_{a}\to1}$ to the instant evaporation result to
get an analytical result that approximates to the actual $f$ over
the entire range covering both low-and high-energy regimes. However,
the simple addition would imply double counting of the contributions
to the intermediate-energy regime. 

To further refine the analytical calculation, we subtract a term proportional
to $e^{-x}$ from Eq.~\eqref{eq:-60} and then add them up. Note that
subtracting such a term does not modify the IR behavior of Eq.~\eqref{eq:-60},
nor does it affect the UV behavior of Eq.~\eqref{eq:-23}. Only the
intermediate-energy regime is reduced. The coefficient of the $e^{-x}$
term, corresponding to $\kappa_{\pm}$ and $\kappa_{0}$ in Eq.~\eqref{eq:-32},
can be determined by the total number of particles $N_{i}$ in Eq.~\eqref{eq:Ni}. 

Assembling the above pieces together, we obtain Eqs.~\eqref{eq:-30}-\eqref{eq:-32}.

\subsection{Heavy particle decay}

Consider a generic process $i\to j+k$ where $i$ is a heavy particle
with a mass $m_{i}$ while $j$ and $k$ are two generic particles
with negligible masses compared to $m_{i}$.  If $i$ decays non-relativistically
in the early universe, the corresponding $\Gamma_{{\rm prod}}$ for
$j$ is approximately a delta function according to Eq.~\eqref{eq:-3}.
In this scenario, Eq.~\eqref{eq:-17} gives
\begin{align}
f_{j} & \approx\int_{0}^{a}\frac{\pi|{\cal M}|^{2}}{2m_{i}^{3}}\frac{n_{i}(a')}{H(a')a'}\delta\left(\frac{ap}{a'}-\frac{m_{i}}{2}\right)da'\nonumber \\
 & \approx\frac{\pi|{\cal M}|^{2}}{m_{i}^{4}}\frac{n_{i}(\tilde{a})}{H(\tilde{a})}\thinspace,\label{eq:-9}
\end{align}
where $|{\cal M}|^{2}$ is the squared amplitude of the decay process
and $\tilde{a}=2ap/m_{i}$.   

In the RD era, assuming $i$ decays non-relativistically at a certain
rate $\Gamma_{iD}$, $n_{i}$ and $H$ as functions of $a$ vary as
follows
\begin{equation}
H(a)=H_{\star}\frac{a_{\star}^{2}}{a^{2}}\thinspace,\ \ n_{i}(a)=n_{i\star}\frac{a_{\star}^{3}}{a^{3}}\exp\left[-\frac{\Gamma_{iD}}{2}\cdot\left(\frac{1}{H(a)}-\frac{1}{H_{\star}}\right)\right],\ \ (\text{RD})\thinspace,\label{eq:-10}
\end{equation}
where the subscript $\star$ denotes a pivotal point which can be
set arbitrarily at any moment when the above approximations are still
valid.  The exponential part in Eq.~\eqref{eq:} comes from $e^{-\Gamma_{iD}t}$
with $t\approx1/(2H)$.  Substituting Eq.~\eqref{eq:-10} into Eq.~\eqref{eq:-9},
we obtain
\begin{equation}
f_{j}\approx\frac{\pi|{\cal M}|^{2}a_{\star}n_{i\star}}{2apH_{\star}m_{i}^{3}}\exp\left\{ -\frac{\Gamma_{iD}}{2H_{\star}}\left[\left(\frac{2ap}{a_{\star}m_{i}}\right)^{2}-1\right]\right\} ,\ \ (\text{RD})\thinspace.\label{eq:-11}
\end{equation}
If we set the pivotal point ``$\star$'' well before $i$ starts
to decay, the $-\frac{1}{H_{\star}}$ term in the exponential function
can be neglected. 

In the matter-dominated (MD) era, Eq.~\eqref{eq:-10} should be changed
to
\begin{equation}
H(a)=H_{\star}\frac{a_{\star}^{3/2}}{a^{3/2}}\thinspace,\ \ n_{i}(a)=n_{i\star}\frac{a_{\star}^{3}}{a^{3}}\exp\left[-\frac{2\Gamma_{iD}}{3}\cdot\left(\frac{1}{H(a)}-\frac{1}{H_{\star}}\right)\right],\ \ (\text{MD})\thinspace,\label{eq:-10-1}
\end{equation}
where we have used $t\approx2/(3H)$. Substituting Eq.~\eqref{eq:-10-1}
into Eq.~\eqref{eq:-9}, we obtain
\begin{equation}
f_{j}\approx\frac{\pi|{\cal M}|^{2}n_{i\star}}{H_{\star}m_{i}^{5/2}}\left(\frac{a_{\star}}{2ap}\right)^{3/2}\exp\left\{ -\frac{2\Gamma_{iD}}{3H_{\star}}\left[\left(\frac{2ap}{a_{\star}m_{i}}\right)^{\frac{3}{2}}-1\right]\right\} ,\ \ (\text{MD})\thinspace.\label{eq:-18}
\end{equation}

It is noteworthy that integrating Eq.~\eqref{eq:-11} or \eqref{eq:-18}
over $p$ yields
\begin{equation}
n_{j}=\int_{p_{\min}}^{p_{\max}}f_{j}(p)\frac{4\pi p^{2}dp}{(2\pi)^{3}}=\frac{n_{i\star}a_{\star}^{3}-n_{i}a^{3}}{a^{3}}\thinspace,\label{eq:-19}
\end{equation}
where $p_{\max}=m_{i}/2$ and $p_{\min}=p_{\max}a_{\star}/a$. Eq.~\eqref{eq:-19}
offers an important cross check of the above analytical calculation
since its physical meaning is straightforward: the total number of
$j$ particles produced in a comoving volume should be equal to the
total number of $i$ particles depleted in the volume.

\section{Decaying spectrum  \label{sec:decay}}

For a generic decaying species $i$ after PBH evaporation, without
assuming non-relativistic or relativistic decay, the subsequent evolution
of $f_{i}$ can be computed analytically.  This requires solving
Eq.~\eqref{eq:-25} with $\Gamma_{i,{\rm prod}}=0$ and
\begin{equation}
\Gamma_{i,{\rm abs}}=\frac{m_{i}}{E}\Gamma_{iD}\thinspace,\label{eq:-61}
\end{equation}
where $\Gamma_{iD}$ is the decay rate of $i$ at rest. 

With the variable transformation $p\rightarrow\tilde{p}=a(t)p$ and
$\tilde{f}(t,\tilde{p})=f(t,p)$, we can rewrite Eq.~\eqref{eq:-25}
as
\begin{equation}
\frac{\partial\tilde{f}_{i}}{\partial t}=-\frac{m_{i}}{E}\Gamma_{iD}\tilde{f}_{i}\thinspace.\label{eq:-62}
\end{equation}
 Eq.~\eqref{eq:-62} should be viewed as a differential equation of
$t$ and $\tilde{p}$. Hence $E=\sqrt{p^{2}+m_{i}^{2}}=\sqrt{\tilde{p}^{2}/a^{2}+m_{i}^{2}}$
should be viewed as a function of $\tilde{p}$ (not $p$) and $t$.
Noticing this point, one can see that Eq.~\eqref{eq:-62} can be 
solved by
\begin{equation}
\tilde{f}_{i}(t,\tilde{p})=\tilde{f}_{i}(t_{\text{ev}},\tilde{p})\exp\left[-m_{i}\Gamma_{iD}\int_{t_{t_{\text{ev}}}}^{t}\frac{\text{d}\tau}{\sqrt{\tilde{p}^{2}/a_{\tau}^{2}+m_{i}^{2}}}\right],\label{eq:-63}
\end{equation}
with $a_{\tau}\equiv a(\tau)$.

Transforming it back from $\tilde{f}(t,\tilde{p})$ to $f(t,p)$,
we obtain the solution in Eq.~\eqref{eq:f-decay}. 

During the RD era, we have $a_{t}\propto t^{1/2}$ and $a_{t}/a_{\tau}=(t/\tau)^{1/2}$.
Then the $\tilde{t}$ integral in Eq.~\eqref{eq:f-decay} can be computed
as follows:
\begin{align}
\tilde{t} & \equiv\int_{t_{\text{ev}}}^{t}\frac{m_{i}{\rm d}\tau}{\sqrt{\left(pa_{t}/a_{\tau}\right)^{2}+m_{i}^{2}}}\nonumber \\
 & =\int_{t_{{\rm ev}}}^{t}\frac{m_{i}\text{d}\tau}{\sqrt{\frac{t}{\tau}p^{2}+m_{i}^{2}}}\nonumber \\
 & =t\left[\frac{E}{m_{i}}-\frac{t_{{\rm ev}}}{t}\frac{E^{({\rm ev})}}{m_{i}}+\frac{1}{2}\frac{p^{2}}{m_{i}^{2}}\ln\frac{(E-m_{i})(E^{({\rm ev})}+m_{i})}{(E+m_{i})(E^{({\rm ev})}-m_{i})}\right],\label{eq:-65}
\end{align}
where $E^{({\rm ev})}\equiv\sqrt{m_{i}^{2}+a_{t}^{2}p^{2}/a_{{\rm ev}}^{2}}$.
Expressing the above result in terms of the dimensionless quantities
$x$, $y$ and $\mu$, it is straightforward to get Eq.~\eqref{eq:-37}.

\section{Collision terms\label{sec:collision}}

For a two-body decay process, $1\to2+3$, the production rate in the
collision term \eqref{eq:-16} reads
\begin{equation}
\Gamma_{{\rm prod}}^{(f_{3})}=\frac{1}{2E_{3}}\int d\Pi_{1}d\Pi_{2}f_{1}\left(1\pm f_{2}\right)\left(2\pi\delta\right)^{4}|{\cal M}|^{2}\thinspace,\label{eq:-1}
\end{equation}
where $d\Pi_{i}\equiv\frac{d^{3}\mathbf{p}_{i}}{(2\pi)^{3}2E_{i}}$
with the subscript $i$ indicating quantities for the $i$-th particles
of the process, $\left(2\pi\delta\right)^{4}$ denotes the Dirac delta
function responsible for momentum conservation, and $|{\cal M}|^{2}$
is the squared amplitude of the process. The $1\pm f_{2}$ in Eq.~\eqref{eq:-1}
accounts for the Bose enhancement or Pauli blocking effect. If this
is neglected and the final states are massless,  Eq.~\eqref{eq:-1}
can be reduced to the following integral~\cite{Li:2022bpp}:
\begin{equation}
\Gamma_{{\rm prod}}^{(f_{3})}\approx\frac{|{\cal M}|^{2}}{2E_{3}^{2}}\int_{p_{1}^{{\rm min}}}^{\infty}dp_{1}\frac{f_{1}p_{1}}{8\pi E_{1}}\thinspace,\label{eq:-2}
\end{equation}
where 
\begin{equation}
p_{1}^{{\rm min}}\equiv\frac{|m_{1}^{2}-4E_{3}^{2}|}{4E_{3}}\thinspace.\label{eq:-4}
\end{equation}
Here $|{\cal M}|^{2}$ is  constant for two-body decay processes,
allowing it to be extracted out of the integral. Note that the absolute
value notation ``$|\ |$'' in Eq.~\eqref{eq:-4} should not be omitted
because $E_{3}$ may exceed $m_{1}/2$ significantly for relativistic
$f_{1}$. 

If $f_{1}$ is a non-relativistic distribution, Eq.~\eqref{eq:-2}
gives
\begin{equation}
\Gamma_{{\rm prod}}^{(f_{3})}\approx\frac{\pi|{\cal M}|^{2}}{2m_{1}^{3}}n_{1}\delta\left(E_{3}-\frac{m_{1}}{2}\right)\thinspace,\label{eq:-3}
\end{equation}
where $n_{1}=\int f_{1}\frac{d^{3}\mathbf{p}_{1}}{(2\pi)^{3}}$ is
the number density of particle 1. Eq.~\eqref{eq:-3} can be obtained
by substituting any analytically simple expressions for $f_{1}$ in
Eq.~\eqref{eq:-2}. One example for such expression is $f_{1}=\lim_{\epsilon\to0^{+}}\frac{8\pi^{2}n_{1}\delta(p_{1}-\epsilon)}{(p_{1}+\epsilon)^{2}}$,
which can be used in Eq.~\eqref{eq:-2} to straightforwardly obtain
Eq.~\eqref{eq:-3}.

If $f_{1}$ approximates to the Maxwell-Boltzmann distribution, i.e.~$f_{1}\propto\text{\ensuremath{e^{-E_{1}/T_{1}}}}$,
Eq.~\eqref{eq:-2} gives
\begin{equation}
\Gamma_{{\rm prod}}^{(f_{3})}\approx\frac{|{\cal M}|^{2}T_{1}}{16\pi E_{3}^{2}}c_{1}\exp\left(-\frac{m_{1}^{2}+4E_{3}^{2}}{4T_{1}E_{3}}\right)\thinspace,\label{eq:-7}
\end{equation}
where $c_{1}\equiv f_{1}/\text{\ensuremath{e^{-E_{1}/T_{1}}}}$ is
a normalization factor that can be determined by the number density:
\begin{equation}
c_{1}=\frac{2\pi^{2}n_{1}}{m_{1}^{2}TK_{2}\left(m_{1}/T_{1}\right)}\thinspace.\label{eq:-8}
\end{equation}
Taking the non-relativistic limit, $T_{1}\ll m_{1}$, one can check
that Eq.~\eqref{eq:-7} reduces to Eq.~\eqref{eq:-3}.

\bibliographystyle{JHEP}
\bibliography{ref}

\providecommand{\href}[2]{#2}\begingroup\raggedright\begin{thebibliography}{10}

\bibitem{Fermi:1949ee}
E.~Fermi, {\it {On the Origin of the Cosmic Radiation}},  {\em Phys. Rev.} {\bf
  75} (1949) 1169--1174.

\bibitem{Blandford:2018iot}
R.~Blandford, D.~Meier, and A.~Readhead, {\it {Relativistic Jets from Active
  Galactic Nuclei}},  {\em Ann. Rev. Astron. Astrophys.} {\bf 57} (2019)
  467--509, [\href{http://www.arxiv.org/abs/1812.06025}{{\tt 1812.06025}}].

\bibitem{Blasi:2001hr}
P.~Blasi, R.~Dick, and E.~W. Kolb, {\it {Ultra-High Energy Cosmic Rays from
  Annihilation of Superheavy Dark Matter}},  {\em Astropart. Phys.} {\bf 18}
  (2002) 57--66, [\href{http://www.arxiv.org/abs/astro-ph/0105232}{{\tt
  astro-ph/0105232}}].

\bibitem{Hill:1982iq}
C.~T. Hill, {\it {Monopolonium}},  {\em Nucl. Phys. B} {\bf 224} (1983)
  469--490.

\bibitem{Birkel:1998nx}
M.~Birkel and S.~Sarkar, {\it {Extremely high-energy cosmic rays from relic
  particle decays}},  {\em Astropart. Phys.} {\bf 9} (1998) 297--309,
  [\href{http://www.arxiv.org/abs/hep-ph/9804285}{{\tt hep-ph/9804285}}].

\bibitem{Kuzmin:1998uv}
V.~Kuzmin and I.~Tkachev, {\it {Ultrahigh-energy cosmic rays, superheavy long
  living particles, and matter creation after inflation}},  {\em JETP Lett.}
  {\bf 68} (1998) 271--275,
  [\href{http://www.arxiv.org/abs/hep-ph/9802304}{{\tt hep-ph/9802304}}].

\bibitem{Bhattacharjee:1999mup}
P.~Bhattacharjee and G.~Sigl, {\it {Origin and propagation of extremely
  high-energy cosmic rays}},  {\em Phys. Rept.} {\bf 327} (2000) 109--247,
  [\href{http://www.arxiv.org/abs/astro-ph/9811011}{{\tt astro-ph/9811011}}].

\bibitem{Torres:2004hk}
D.~F. Torres and L.~A. Anchordoqui, {\it {Astrophysical origins of ultrahigh
  energy cosmic rays}},  {\em Rept. Prog. Phys.} {\bf 67} (2004) 1663--1730,
  [\href{http://www.arxiv.org/abs/astro-ph/0402371}{{\tt astro-ph/0402371}}].

\bibitem{Anchordoqui:2018qom}
L.~A. Anchordoqui, {\it {Ultra-High-Energy Cosmic Rays}},  {\em Phys. Rept.}
  {\bf 801} (2019) 1--93, [\href{http://www.arxiv.org/abs/1807.09645}{{\tt
  1807.09645}}].

\bibitem{Ackermann:2022rqc}
M.~Ackermann {\em et~al.}, {\it {High-energy and ultra-high-energy neutrinos: A
  Snowmass white paper}},  {\em JHEAp} {\bf 36} (2022) 55--110,
  [\href{http://www.arxiv.org/abs/2203.08096}{{\tt 2203.08096}}].

\bibitem{Carr:2009jm}
B.~J. Carr, K.~Kohri, Y.~Sendouda, and J.~Yokoyama, {\it {New cosmological
  constraints on primordial black holes}},  {\em Phys. Rev. D} {\bf 81} (2010)
  104019, [\href{http://www.arxiv.org/abs/0912.5297}{{\tt 0912.5297}}].

\bibitem{Carr:2020gox}
B.~Carr, K.~Kohri, Y.~Sendouda, and J.~Yokoyama, {\it {Constraints on
  primordial black holes}},  {\em Rept. Prog. Phys.} {\bf 84} (2021), no.~11
  116902, [\href{http://www.arxiv.org/abs/2002.12778}{{\tt 2002.12778}}].

\bibitem{Lunardini:2019zob}
C.~Lunardini and Y.~F. Perez-Gonzalez, {\it {Dirac and Majorana neutrino
  signatures of primordial black holes}},  {\em JCAP} {\bf 08} (2020) 014,
  [\href{http://www.arxiv.org/abs/1910.07864}{{\tt 1910.07864}}].

\bibitem{Halzen:1995hu}
F.~Halzen, B.~Keszthelyi, and E.~Zas, {\it {Neutrinos from primordial black
  holes}},  {\em Phys. Rev. D} {\bf 52} (1995) 3239--3247,
  [\href{http://www.arxiv.org/abs/hep-ph/9502268}{{\tt hep-ph/9502268}}].

\bibitem{Bugaev:2000bz}
E.~V. Bugaev and K.~V. Konishchev, {\it {Constraints on diffuse neutrino
  background from primordial black holes}},  {\em Phys. Rev. D} {\bf 65} (2002)
  123005, [\href{http://www.arxiv.org/abs/astro-ph/0005295}{{\tt
  astro-ph/0005295}}].

\bibitem{Bugaev:2002yt}
E.~V. Bugaev and K.~V. Konishchev, {\it {Cosmological constraints from
  evaporations of primordial black holes}},  {\em Phys. Rev. D} {\bf 66} (2002)
  084004, [\href{http://www.arxiv.org/abs/astro-ph/0206082}{{\tt
  astro-ph/0206082}}].

\bibitem{Wang:2020uvi}
S.~Wang, D.-M. Xia, X.~Zhang, S.~Zhou, and Z.~Chang, {\it {Constraining
  primordial black holes as dark matter at JUNO}},  {\em Phys. Rev. D} {\bf
  103} (2021), no.~4 043010, [\href{http://www.arxiv.org/abs/2010.16053}{{\tt
  2010.16053}}].

\bibitem{DeRomeri:2021xgy}
V.~De~Romeri, P.~Mart\'\i{}nez-Mirav\'e, and M.~T\'ortola, {\it {Signatures of
  primordial black hole dark matter at DUNE and THEIA}},  {\em JCAP} {\bf 10}
  (2021) 051, [\href{http://www.arxiv.org/abs/2106.05013}{{\tt 2106.05013}}].

\bibitem{Calabrese:2021zfq}
R.~Calabrese, D.~F.~G. Fiorillo, G.~Miele, S.~Morisi, and A.~Palazzo, {\it
  {Primordial black hole dark matter evaporating on the neutrino floor}},  {\em
  Phys. Lett. B} {\bf 829} (2022) 137050,
  [\href{http://www.arxiv.org/abs/2106.02492}{{\tt 2106.02492}}].

\bibitem{Bernal:2022swt}
N.~Bernal, V.~Mu\~noz Albornoz, S.~Palomares-Ruiz, and P.~Villanueva-Domingo,
  {\it {Current and future neutrino limits on the abundance of primordial black
  holes}},  {\em JCAP} {\bf 10} (2022) 068,
  [\href{http://www.arxiv.org/abs/2203.14979}{{\tt 2203.14979}}].

\bibitem{Liu:2023cqs}
Q.~Liu and K.~C.~Y. Ng, {\it {The Sensitivity Floor for Primordial Black Holes
  with Neutrino Searches}},  \href{http://www.arxiv.org/abs/2312.06108}{{\tt
  2312.06108}}.

\bibitem{DeRomeri:2024zqs}
V.~De~Romeri, Y.~F. Perez-Gonzalez, and A.~Tolino, {\it {Primordial black hole
  probes of heavy neutral leptons}},
  \href{http://www.arxiv.org/abs/2405.00124}{{\tt 2405.00124}}.

\bibitem{Baldes:2020nuv}
I.~Baldes, Q.~Decant, D.~C. Hooper, and L.~Lopez-Honorez, {\it {Non-Cold Dark
  Matter from Primordial Black Hole Evaporation}},  {\em JCAP} {\bf 08} (2020)
  045, [\href{http://www.arxiv.org/abs/2004.14773}{{\tt 2004.14773}}].

\bibitem{Gondolo:2020uqv}
P.~Gondolo, P.~Sandick, and B.~Shams Es~Haghi, {\it {Effects of primordial
  black holes on dark matter models}},  {\em Phys. Rev. D} {\bf 102} (2020),
  no.~9 095018, [\href{http://www.arxiv.org/abs/2009.02424}{{\tt 2009.02424}}].

\bibitem{Bernal:2020kse}
N.~Bernal and O.~Zapata, {\it {Self-interacting Dark Matter from Primordial
  Black Holes}},  {\em JCAP} {\bf 03} (2021) 007,
  [\href{http://www.arxiv.org/abs/2010.09725}{{\tt 2010.09725}}].

\bibitem{Bernal:2020bjf}
N.~Bernal and O.~Zapata, {\it {Dark Matter in the Time of Primordial Black
  Holes}},  {\em JCAP} {\bf 03} (2021) 015,
  [\href{http://www.arxiv.org/abs/2011.12306}{{\tt 2011.12306}}].

\bibitem{Bernal:2020ili}
N.~Bernal and O.~Zapata, {\it {Gravitational dark matter production: primordial
  black holes and UV freeze-in}},  {\em Phys. Lett. B} {\bf 815} (2021) 136129,
  [\href{http://www.arxiv.org/abs/2011.02510}{{\tt 2011.02510}}].

\bibitem{Auffinger:2020afu}
J.~Auffinger, I.~Masina, and G.~Orlando, {\it {Bounds on warm dark matter from
  Schwarzschild primordial black holes}},  {\em Eur. Phys. J. Plus} {\bf 136}
  (2021), no.~2 261, [\href{http://www.arxiv.org/abs/2012.09867}{{\tt
  2012.09867}}].

\bibitem{Hooper:2020otu}
D.~Hooper and G.~Krnjaic, {\it {GUT Baryogenesis With Primordial Black Holes}},
   {\em Phys. Rev. D} {\bf 103} (2021), no.~4 043504,
  [\href{http://www.arxiv.org/abs/2010.01134}{{\tt 2010.01134}}].

\bibitem{Datta:2020bht}
S.~Datta, A.~Ghosal, and R.~Samanta, {\it {Baryogenesis from ultralight
  primordial black holes and strong gravitational waves from cosmic strings}},
  {\em JCAP} {\bf 08} (2021) 021,
  [\href{http://www.arxiv.org/abs/2012.14981}{{\tt 2012.14981}}].

\bibitem{Cheek:2021odj}
A.~Cheek, L.~Heurtier, Y.~F. Perez-Gonzalez, and J.~Turner, {\it {Primordial
  black hole evaporation and dark matter production. I. Solely Hawking
  radiation}},  {\em Phys. Rev. D} {\bf 105} (2022), no.~1 015022,
  [\href{http://www.arxiv.org/abs/2107.00013}{{\tt 2107.00013}}].

\bibitem{Masina:2021zpu}
I.~Masina, {\it {Dark Matter and Dark Radiation from Evaporating Kerr
  Primordial Black Holes}},  {\em Grav. Cosmol.} {\bf 27} (2021), no.~4
  315--330, [\href{http://www.arxiv.org/abs/2103.13825}{{\tt 2103.13825}}].

\bibitem{Cheek:2021cfe}
A.~Cheek, L.~Heurtier, Y.~F. Perez-Gonzalez, and J.~Turner, {\it {Primordial
  black hole evaporation and dark matter production. II. Interplay with the
  freeze-in or freeze-out mechanism}},  {\em Phys. Rev. D} {\bf 105} (2022),
  no.~1 015023, [\href{http://www.arxiv.org/abs/2107.00016}{{\tt 2107.00016}}].

\bibitem{Sandick:2021gew}
P.~Sandick, B.~S. Es~Haghi, and K.~Sinha, {\it {Asymmetric reheating by
  primordial black holes}},  {\em Phys. Rev. D} {\bf 104} (2021), no.~8 083523,
  [\href{http://www.arxiv.org/abs/2108.08329}{{\tt 2108.08329}}].

\bibitem{Bernal:2021yyb}
N.~Bernal, F.~Hajkarim, and Y.~Xu, {\it {Axion Dark Matter in the Time of
  Primordial Black Holes}},  {\em Phys. Rev. D} {\bf 104} (2021) 075007,
  [\href{http://www.arxiv.org/abs/2107.13575}{{\tt 2107.13575}}].

\bibitem{Bernal:2021bbv}
N.~Bernal, Y.~F. Perez-Gonzalez, Y.~Xu, and O.~Zapata, {\it {ALP dark matter in
  a primordial black hole dominated universe}},  {\em Phys. Rev. D} {\bf 104}
  (2021), no.~12 123536, [\href{http://www.arxiv.org/abs/2110.04312}{{\tt
  2110.04312}}].

\bibitem{Calabrese:2021src}
R.~Calabrese, M.~Chianese, D.~F.~G. Fiorillo, and N.~Saviano, {\it {Direct
  detection of light dark matter from evaporating primordial black holes}},
  {\em Phys. Rev. D} {\bf 105} (2022), no.~2 L021302,
  [\href{http://www.arxiv.org/abs/2107.13001}{{\tt 2107.13001}}].

\bibitem{JyotiDas:2021shi}
S.~Jyoti~Das, D.~Mahanta, and D.~Borah, {\it {Low scale leptogenesis and dark
  matter in the presence of primordial black holes}},  {\em JCAP} {\bf 11}
  (2021) 019, [\href{http://www.arxiv.org/abs/2104.14496}{{\tt 2104.14496}}].

\bibitem{Bernal:2022pue}
N.~Bernal, C.~S. Fong, Y.~F. Perez-Gonzalez, and J.~Turner, {\it {Rescuing
  high-scale leptogenesis using primordial black holes}},  {\em Phys. Rev. D}
  {\bf 106} (2022), no.~3 035019,
  [\href{http://www.arxiv.org/abs/2203.08823}{{\tt 2203.08823}}].

\bibitem{Bernal:2022oha}
N.~Bernal, Y.~F. Perez-Gonzalez, and Y.~Xu, {\it {Superradiant production of
  heavy dark matter from primordial black holes}},  {\em Phys. Rev. D} {\bf
  106} (2022), no.~1 015020, [\href{http://www.arxiv.org/abs/2205.11522}{{\tt
  2205.11522}}].

\bibitem{Coleppa:2022pnf}
B.~Coleppa, K.~Loho, and S.~Shil, {\it {Dark Sector extensions of the Littlest
  Seesaw in the presence of Primordial Black Holes}},  {\em JCAP} {\bf 06}
  (2023) 027, [\href{http://www.arxiv.org/abs/2209.06793}{{\tt 2209.06793}}].

\bibitem{Gehrman:2022imk}
T.~C. Gehrman, B.~Shams Es~Haghi, K.~Sinha, and T.~Xu, {\it {Baryogenesis,
  primordial black holes and MHz\textendash{}GHz gravitational waves}},  {\em
  JCAP} {\bf 02} (2023) 062, [\href{http://www.arxiv.org/abs/2211.08431}{{\tt
  2211.08431}}].

\bibitem{Calabrese:2023key}
R.~Calabrese, M.~Chianese, J.~Gunn, G.~Miele, S.~Morisi, and N.~Saviano, {\it
  {Limits on light primordial black holes from high-scale leptogenesis}},  {\em
  Phys. Rev. D} {\bf 107} (2023), no.~12 123537,
  [\href{http://www.arxiv.org/abs/2305.13369}{{\tt 2305.13369}}].

\bibitem{Schmitz:2023pfy}
K.~Schmitz and X.-J. Xu, {\it {Wash-in leptogenesis after the evaporation of
  primordial black holes}},  {\em Phys. Lett. B} {\bf 849} (2024) 138473,
  [\href{http://www.arxiv.org/abs/2311.01089}{{\tt 2311.01089}}].

\bibitem{Gehrman:2023esa}
T.~C. Gehrman, B.~Shams Es~Haghi, K.~Sinha, and T.~Xu, {\it {The primordial
  black holes that disappeared: connections to dark matter and MHz-GHz
  gravitational Waves}},  {\em JCAP} {\bf 10} (2023) 001,
  [\href{http://www.arxiv.org/abs/2304.09194}{{\tt 2304.09194}}].

\bibitem{Gehrman:2023qjn}
T.~C. Gehrman, B.~Shams Es~Haghi, K.~Sinha, and T.~Xu, {\it {Recycled dark
  matter}},  {\em JCAP} {\bf 03} (2024) 044,
  [\href{http://www.arxiv.org/abs/2310.08526}{{\tt 2310.08526}}].

\bibitem{Arcadi:2024tib}
G.~Arcadi, M.~Lindner, J.~P. Neto, and F.~S. Queiroz, {\it {Ultraheavy Dark
  Matter and WIMPs Production aided by Primordial Black Holes}},
  \href{http://www.arxiv.org/abs/2408.13313}{{\tt 2408.13313}}.

\bibitem{IceCube-Gen2:2020qha}
{\bf IceCube-Gen2} {\bf Collaboration}, M.~G. Aartsen {\em et~al.}, {\it
  {IceCube-Gen2: the window to the extreme Universe}},  {\em J. Phys. G} {\bf
  48} (2021), no.~6 060501, [\href{http://www.arxiv.org/abs/2008.04323}{{\tt
  2008.04323}}].

\bibitem{GRAND:2018iaj}
{\bf GRAND} {\bf Collaboration}, J.~\'Alvarez-Mu\~niz {\em et~al.}, {\it {The
  Giant Radio Array for Neutrino Detection (GRAND): Science and Design}},  {\em
  Sci. China Phys. Mech. Astron.} {\bf 63} (2020), no.~1 219501,
  [\href{http://www.arxiv.org/abs/1810.09994}{{\tt 1810.09994}}].

\bibitem{Page:1976df}
D.~N. Page, {\it {Particle Emission Rates from a Black Hole: Massless Particles
  from an Uncharged, Nonrotating Hole}},  {\em Phys. Rev. D} {\bf 13} (1976)
  198--206.

\bibitem{Carr:1975qj}
B.~J. Carr, {\it {The Primordial black hole mass spectrum}},  {\em Astrophys.
  J.} {\bf 201} (1975) 1--19.

\bibitem{Keith:2020jww}
C.~Keith, D.~Hooper, N.~Blinov, and S.~D. McDermott, {\it {Constraints on
  Primordial Black Holes From Big Bang Nucleosynthesis Revisited}},  {\em Phys.
  Rev. D} {\bf 102} (2020), no.~10 103512,
  [\href{http://www.arxiv.org/abs/2006.03608}{{\tt 2006.03608}}].

\bibitem{Vitagliano:2019yzm}
E.~Vitagliano, I.~Tamborra, and G.~Raffelt, {\it {Grand Unified Neutrino
  Spectrum at Earth: Sources and Spectral Components}},  {\em Rev. Mod. Phys.}
  {\bf 92} (2020) 45006, [\href{http://www.arxiv.org/abs/1910.11878}{{\tt
  1910.11878}}].

\bibitem{MacGibbon:1991vc}
J.~H. MacGibbon and B.~J. Carr, {\it {Cosmic rays from primordial black
  holes}},  {\em Astrophys. J.} {\bf 371} (1991) 447--469.

\bibitem{Capanema:2021hnm}
A.~Capanema, A.~Esmaeili, and A.~Esmaili, {\it {Evaporating primordial black
  holes in gamma ray and neutrino telescopes}},  {\em JCAP} {\bf 12} (2021),
  no.~12 051, [\href{http://www.arxiv.org/abs/2110.05637}{{\tt 2110.05637}}].

\bibitem{Poulin:2016anj}
V.~Poulin, J.~Lesgourgues, and P.~D. Serpico, {\it {Cosmological constraints on
  exotic injection of electromagnetic energy}},  {\em JCAP} {\bf 03} (2017)
  043, [\href{http://www.arxiv.org/abs/1610.10051}{{\tt 1610.10051}}].

\bibitem{Xu:2020qek}
X.-J. Xu, {\it {The $\nu_{R}$-philic scalar: its loop-induced interactions and
  Yukawa forces in LIGO observations}},  {\em JHEP} {\bf 09} (2020) 105,
  [\href{http://www.arxiv.org/abs/2007.01893}{{\tt 2007.01893}}].

\bibitem{Chauhan:2020mgv}
G.~Chauhan and X.-J. Xu, {\it {How dark is the $\nu_R$-philic dark photon?}},
  {\em JHEP} {\bf 04} (2021) 003,
  [\href{http://www.arxiv.org/abs/2012.09980}{{\tt 2012.09980}}].

\bibitem{Xu:2023xva}
X.-J. Xu, S.~Zhou, and J.~Zhu, {\it {The $\nu_{R}$-philic scalar dark matter}},
   {\em JCAP} {\bf 04} (2024) 012,
  [\href{http://www.arxiv.org/abs/2310.16346}{{\tt 2310.16346}}].

\bibitem{IceCube:2021uhz}
{\bf IceCube} {\bf Collaboration}, R.~Abbasi {\em et~al.}, {\it {Improved
  Characterization of the Astrophysical Muon\textendash{}neutrino Flux with 9.5
  Years of IceCube Data}},  {\em Astrophys. J.} {\bf 928} (2022), no.~1 50,
  [\href{http://www.arxiv.org/abs/2111.10299}{{\tt 2111.10299}}].

\bibitem{ARA:2019wcf}
{\bf ARA} {\bf Collaboration}, P.~Allison {\em et~al.}, {\it {Constraints on
  the diffuse flux of ultrahigh energy neutrinos from four years of Askaryan
  Radio Array data in two stations}},  {\em Phys. Rev. D} {\bf 102} (2020),
  no.~4 043021, [\href{http://www.arxiv.org/abs/1912.00987}{{\tt 1912.00987}}].

\bibitem{Anker:2019rzo}
A.~Anker {\em et~al.}, {\it {A search for cosmogenic neutrinos with the ARIANNA
  test bed using 4.5 years of data}},  {\em JCAP} {\bf 03} (2020) 053,
  [\href{http://www.arxiv.org/abs/1909.00840}{{\tt 1909.00840}}].

\bibitem{PierreAuger:2019ens}
{\bf Pierre Auger} {\bf Collaboration}, A.~Aab {\em et~al.}, {\it {Probing the
  origin of ultra-high-energy cosmic rays with neutrinos in the EeV energy
  range using the Pierre Auger Observatory}},  {\em JCAP} {\bf 10} (2019) 022,
  [\href{http://www.arxiv.org/abs/1906.07422}{{\tt 1906.07422}}].

\bibitem{KM3Net:2016zxf}
{\bf KM3Net} {\bf Collaboration}, S.~Adrian-Martinez {\em et~al.}, {\it {Letter
  of intent for KM3NeT 2.0}},  {\em J. Phys. G} {\bf 43} (2016), no.~8 084001,
  [\href{http://www.arxiv.org/abs/1601.07459}{{\tt 1601.07459}}].

\bibitem{Wissel:2020sec}
S.~Wissel {\em et~al.}, {\it {Prospects for high-elevation radio detection of
  \ensuremath{>}100 PeV tau neutrinos}},  {\em JCAP} {\bf 11} (2020) 065,
  [\href{http://www.arxiv.org/abs/2004.12718}{{\tt 2004.12718}}].

\bibitem{Prohira:2019glh}
S.~Prohira {\em et~al.}, {\it {Observation of Radar Echoes From High-Energy
  Particle Cascades}},  {\em Phys. Rev. Lett.} {\bf 124} (2020), no.~9 091101,
  [\href{http://www.arxiv.org/abs/1910.12830}{{\tt 1910.12830}}].

\bibitem{Romero-Wolf:2020pzh}
A.~Romero-Wolf {\em et~al.}, {\it {An Andean Deep-Valley Detector for
  High-Energy Tau Neutrinos}},  in {\em {Latin American Strategy Forum for
  Research Infrastructure}}, 2, 2020.
\newblock \href{http://www.arxiv.org/abs/2002.06475}{{\tt 2002.06475}}.

\bibitem{Otte:2019aaf}
A.~N. Otte, A.~M. Brown, M.~Doro, A.~Falcone, J.~Holder, E.~Judd, P.~Kaaret,
  M.~Mariotti, K.~Murase, and I.~Taboada, {\it {Trinity: An Air-Shower Imaging
  Instrument to detect Ultrahigh Energy Neutrinos}},
  \href{http://www.arxiv.org/abs/1907.08727}{{\tt 1907.08727}}.

\bibitem{Li:2022bpp}
S.-P. Li and X.-J. Xu, {\it {Dark matter produced from right-handed
  neutrinos}},  {\em JCAP} {\bf 06} (2023) 047,
  [\href{http://www.arxiv.org/abs/2212.09109}{{\tt 2212.09109}}].

\end{thebibliography}\endgroup

\end{document}